\documentclass[namedreferences]{kluwer}

\usepackage{rotating}
\usepackage{graphicx}
\usepackage{psfrag}

\def\lta{~\raise.4ex\hbox{$<$}\llap{\lower.6ex\hbox{$\sim$}}~}
\def\gta{~\raise.4ex\hbox{$>$}\llap{\lower.6ex\hbox{$\sim$}}~}

\def\ie{{\it i.e.}}
\def\eg{{\it e.g.}}

\newcommand{\aap}{Astron. \& Astrophys.}

\newcommand{\apj}{Astrophys. J.}
\newcommand{\aj}{Astron. J.}

\newcommand{\mnras}{Mon. Not. R. Astron. Soc.}

\begin{document}
\begin{article}

\begin{opening}\title{Injection of Oort Cloud Comets: The Fundamental Role of Stellar Perturbations}

\author{Hans Rickman \email{hans@astro.uu.se}}
\institute{PAN Space Research Center, Bartycka 18A, PL-00-716, Warszawa, Poland\\
            Uppsala Astronomical Observatory, Box 515, SE-75120 Uppsala, Sweden}

\author{Marc Fouchard \email{fouchard@imcce.fr}}
\institute{LAL-IMCCE/USTL, 1 Impasse de l'observatoire, F-59000 Lille, France}

\author{Christiane Froeschl\'e \email{froesch@obs-nice.fr}}
\institute{Observatoire de la C\^ote d'Azur, UMR 6202, Bv. de l'Observatoire, B.P. 4229, F-06304 Nice cedex 4, France}

\author{Giovanni B. Valsecchi \email{giovanni@iasf-roma.inaf.it}}
\institute{INAF-IASF, via Fosso del Cavaliere 100, I-00133 Roma, Italy}

\begin{abstract}
We present Monte Carlo simulations of the dynamical evolution of the Oort cloud over the age of the 
Solar System, using an initial sample of one million test comets without any cloning. Our model 
includes perturbations due to the Galactic tide (radial and vertical) and passing stars. We present 
the first detailed analysis of the injection mechanism into observable orbits by comparing the 
complete model with separate models for tidal and stellar perturbations alone. We find that a 
fundamental role for injecting comets from the region outside the loss cone (perihelion distance 
$q > 15$~AU) into observable orbits ($q < 5$~AU) is played by stellar perturbations. These act in 
synergy with the tide such that the total injection rate is significantly larger than the sum of 
the two separate rates. This synergy is as important during comet showers as during quiescent periods 
and concerns comets with both small and large semi-major axes. We propose different dynamical 
mechanisms to explain the synergies in the inner and outer parts of the Oort Cloud. We find that 
the filling of the observable part of the loss cone under normal conditions in the present-day 
Solar System rises from $<1$\% for $a < 20\,000$~AU to about 100\% for $a \gta 100\,000$~AU.
\end{abstract}

\end{opening}

%\newpage
%####################################################################################################
%###################################################################################################

\section{Introduction}

When analyzing the distribution of original inverse semi-major axes of long-period comets,
\inlinecite{OOR:50} concluded that the near-parabolic spike of this distribution reveals a 
distant reservoir of comets (the `Oort Cloud'). His favoured mechanism
of injection of comets from this reservoir into observable orbits (\ie, with small perihelion
distances) was the passage of stars in the vicinity of the reservoir, whereby the long-term
reshuffling of angular momenta would ensure a continuous infeed into the innermost part
of the Solar System.

Until the 1980's stellar perturbation was the only mechanism considered, when issues
concerning the injection of comets from the Oort Cloud were discussed (\eg, \opencite{RIC:76}; 
\opencite{WEI:79}; \opencite{FER:80}; \opencite{HIL:81}; \opencite{REM.MIG:85}). However, by that
time it became clear that the tidal action of the Galaxy as a whole must also have an
important influence -- especially the part corresponding to the $z$-dependent disk
potential \cite{HEI.TRE:86}. This was verified by noting that the Galactic
latitudes of perihelia of new Oort Cloud comets have a double-peaked distribution that is
characteristic of the disk tide \cite{DEL:87}.

An important paper by \inlinecite{DUNetal:87} treated the formation of the Oort Cloud and
showed that the characteristic time scale for changing the perihelion distances, independent 
of the semi-major axis, is shorter for the Galactic disk tide than for the stellar 
perturbations. This has been verified by later analytic work, \eg, by \inlinecite{FER:05}, 
and further numerical simulations of Oort Cloud evolution (\eg, \opencite{HEIetal:87}) have 
given support to the dominance of Galactic tides for comet injection.

Consequently, stellar perturbations have come to be practically neglected as a source of
comet injection -- except when discussing ``comet showers''  \cite{HIL:81} arising from
rare stellar passages through the dense, inner parts of the Oort Cloud.
The importance of stellar perturbations for randomizing
the orbital distribution of the Oort Cloud and thus keeping the relevant infeed trajectories of
the disk tide populated over long time scales has been realized (see \opencite{DYB:02} and 
references therein), 
but the actual injection is often seen as due only to the tide. Hence it should be
limited to semi-major axes large enough for the tidal perturbation to bring the cometary
perihelion at once from outside the ``Jupiter-Saturn barrier'' (\ie, perihelion distance $q
\gta 15$~AU) into the observable region ($q < 5$~AU). The result is that one expects new
comets to have $a_{ori} \gta 3\times 10^4$~AU \cite{BAI.STA:88,LEVetal:01,FER:05}.

On the other hand, some recent papers indicate that this picture may have problems. 
The fractional population of the observable region -- if fed only by Galactic
tides -- is small enough, and the orbital periods long enough, that the estimated total
population of the Oort Cloud may be uncomfortably large
\cite{CHA.MOR:07}. And in addition, when non-gravitational effects
are included into orbit determinations for new Oort Cloud comets \cite{KRO:06},
the resulting original orbits tend to be of shorter periods, having smaller semi-major axes
-- often much smaller than $3\times 10^4$~AU.

Meanwhile, we have developed fast and accurate methods to treat both the Galactic tides
\cite{BREetal:07,FOUetal:07b} and stellar perturbations \cite{RICetal:05}
in Monte Carlo simulations of Oort Cloud dynamics. This has allowed us to perform
calculations, to be presented here, where the cloud is represented by as many as $10^6$
sample comets and integrated over a time exceeding the age of the Solar System.
This amounts to $5\times 10^{15}$ comet-years of individual evolutions (or only 
slightly smaller due to the loss of comets during the simulation), which is 
much more than in all previous long-term Oort cloud simulations -- \eg, $3\times 10^{13}$ 
comet-years for \inlinecite{DUNetal:87}, 
$4\times 10^{14}$ comet-years for \inlinecite{MAZ:06}, and $\sim 4\times 10^{13}$ 
comet-years for \inlinecite{EMEetal:07} who used cloning. \inlinecite{HEI:90} 
simulated $\simeq 7\times 10^{15}$ comet-years but only thanks to extensive cloning 
during the course of the simulation. In fact her long-term simulations (4.5~Gyr)
concerned only $\simeq10^4$ ``tokens'', \ie, comets actually followed, while these 
were meant to represent $\simeq150$ times as many comets by means of cloning.

Our work is the first to study the mechanism of injection of comets from the Oort 
Cloud over the age of the Solar System by simulating and comparing different 
dynamical models. The reason why models involving both the Galactic tide and stellar
perturbations gave a much higher flux of injected comets than those involving only 
stars \cite{HEIetal:87,HEI:90} was never clarified, since comparisons with models involving 
only the tide were not made. In the present paper we concentrate on a comprehensive 
comparison of combined and separate models, thus describing and analyzing for the 
first time the synergy effect of Galactic tide and stars.

We also take special care to define correctly the encounter velocities in our sample 
of passing stars, thereby arriving at somewhat larger values than those used 
previously. Finally, we study the relative filling of the observable part of the 
loss cone and the distribution of inverse semi-major axes of the injected comets. 
These studies are, however, only preliminary, since our current simulations do not 
include planetary perturbations, and thus we cannot account for those comets that 
arrive into the observable region after having ``diffused'' across the Jupiter-Saturn 
barrier in several revolutions.

Our calculations are presented in Sect.~2, and in Sects.~3--5 we describe our results 
in terms of the distribution of injection times into the inner planetary system, the 
flux of new, observable comets as a function of time, and the distributions of inverse 
semi-major axis and Galactic latitude of perihelion as well as loss cone filling at 
representative epochs. In Sect.~6 we discuss the results and summarize our conclusions.

\section{Calculations}

As a simplifying assumption, we consider the Oort Cloud to have been formed
instantaneously at a given epoch, and that its orbital distribution was isotropic to begin
with. Thus the initial conditions are chosen with flat distributions of $\cos i_o$, $\omega_o$,
$\Omega_o$ and $M_o$ (we use common notations for the orbital elements, and the angles
may be defined with respect to an arbitrary frame of reference). We consider a thermalized 
initial state of the cloud, where the semi-major axes ($a_o$) are chosen in the interval 
$3\times 10^3 < a_o < 1 \times 10^5$~AU with a probability density $\propto a_o^{-1.5}$ 
\cite{DUNetal:87}. The initial eccentricities ($e_o$) are chosen with a
density function $\propto e_o$ in such a way that the perihelia are outside the planetary
system ($q>32$~AU). We thus initialise $1\times 10^6$ comets.

The Galactic parameters used for calculating the tidal effects are the same as described 
in earlier papers \cite{FOUetal:07b}. The most important one for comparison with other 
investigations is the mid-plane disk density, which we take as 0.1~$M_\odot$pc$^{-3}$ 
\cite{HOL.FLY:00}. This is in agreement with \inlinecite{EMEetal:07}, while 
\inlinecite{HEI:90} used 0.18~$M_\odot$pc$^{-3}$ \cite{BAH:84}.

The simulation runs with a predefined set of $197\,906$ stellar encounters, occurring at
random times during a lapse of $t_{max}=5\times 10^9$~yr, with random solar impact parameters 
up to $d_{max}=4\times 10^5$~AU, and with random stellar masses and velocities. Our procedure 
for creating each of these encounters is as follows. Let $\xi$ denote a stochastic, random number 
with a uniform probability distribution on the interval $[0,1]$. The solar impact parameter is 
chosen as $d=\xi_d^{1/2}\times d_{max}$, and the time of the encounter (specifically, the time 
of the star's perihelion passage) is $t=\xi_t\times t_{max}$. The direction of stellar 
motion with respect to the Sun is defined in terms of Galactic latitude and longitude ($b,\ell$) 
such that $\sin b=2\xi_b-1$ and $\ell=\xi_\ell\times2\pi$, \ie, it has an isotropic 
distribution. The point at which the initial stellar velocity cuts the orthogonal impact plane 
is situated on a circle of radius $d$ around the Sun, and its location is defined by an 
azimuthal angle ($a$) such that $a=\xi_a\times2\pi$.

Next we choose the type of the star. We use 13 categories as in \inlinecite{RICetal:04} with 
parameters listed in Table~\ref{tab:enc}. To each category we associate one value of the 
stellar mass. These masses are generally taken as those of the archetypal spectral classes 
along the main sequence according to \inlinecite{ALL:85}. However, 
in contrast to our earlier investigations, we replace the archetypal mass of 18~$M_\odot$ 
for B0 stars by a weighted average of 9~$M_\odot$, considering that the less massive, later 
types (B2, B5) are much more common than the earlier ones. The relative encounter frequencies 
$f_i$ of Table~\ref{tab:enc} are taken from \inlinecite{GARetal:01}, where they were derived 
from the respective products of number density and mean velocity, $n_i\langle v_i\rangle$. 
A random number $\xi_i$ is used to pick a stellar category $i$ with the probability 
$f_i/\sum f_i$.

\begin{table}
\begin{tabular}{|c|ccrr|rr|}
Type & Mass & Enc. freq. & $v_\odot$ & $\sigma_\ast$ & $\langle V\rangle$ & $\sigma_V$ \\
~ & ($M_\odot$) & ~ & (km/s) & (km/s) & (km/s) & (km/s) \\
\hline
B0 & 9 & 0.005 & 18.6 & 14.7 & 24.6 & 6.7 \\
A0 & 3.2 & 0.03 & 17.1 & 19.7 & 27.5 & 9.3 \\
A5 & 2.1 & 0.04 & 13.7 & 23.7 & 29.3 & 10.4 \\
F0 & 1.7 & 0.15 & 17.1 & 29.1 & 36.5 & 12.6 \\
F5 & 1.3 & 0.08 & 17.1 & 36.2 & 43.6 & 15.6 \\
G0 & 1.1 & 0.22 & 26.4 & 37.4 & 49.8 & 17.1 \\
G5 & 0.93 & 0.35 & 23.9 & 39.2 & 49.6 & 17.9 \\
K0 & 0.78 & 0.34 & 19.8 & 34.1 & 42.6 & 15.0 \\
K5 & 0.69 & 0.85 & 25.0 & 43.4 & 54.3 & 19.2 \\
M0 & 0.47 & 1.29 & 17.3 & 42.7 & 50.0 & 18.0 \\
M5 & 0.21 & 6.39 & 23.3 & 41.8 & 51.8 & 18.3 \\
wd & 0.9 & 0.72 & 38.3 & 63.4 & 80.2 & 28.2 \\
gi & 4 & 0.06 & 21.0 & 41.0 & 49.7 & 17.5 \\
\hline
\end{tabular}
\caption{Stellar parameters. The types are mostly MK types for main sequence stars; `wd' 
indicates white dwarfs, and `gi' indicates giant stars. The encounter frequencies are given 
in number per Myr within 1 pc. The following two columns list the solar apex velocity with 
respect to the corresponding type, and the spherical Maxwellian velocity dispersion. The 
last two columns give the mean heliocentric encounter velocity and its standard deviation 
according to our results.\vspace{.3cm}
}\label{tab:enc}
\end{table}

Finally, we choose the speed of the stellar motion in the following way. The velocity 
dispersions ($\sigma_{\ast i}$) listed in Table~\ref{tab:enc} are taken from 
\inlinecite{GARetal:01}, and they correspond to the semi-axes of the velocity ellipsoids 
($\sigma_{ui}$, $\sigma_{vi}$, $\sigma_{wi}$) listed by \inlinecite{MIH.BIN:81} using: 
$\sigma_{\ast i}^2=\sigma_{ui}^2+\sigma_{vi}^2+\sigma_{wi}^2$. For the peculiar velocity 
($v_\ast$) of a star with respect to its LSR, we use a spherical Maxwellian by taking 
$\eta_u$, $\eta_v$ and $\eta_w$ as three random numbers, each with a Gaussian probability 
distribution with expectance 0 and variance 1, and computing 
$v_\ast=\sigma_{\ast i}\left\{(\eta_u^2+\eta_v^2+\eta_w^2)/3\right\}^{1/2}$. The star's 
heliocentric velocity is found by combining the vector ${\mathbf v}_\ast$ with the Sun's 
peculiar velocity with respect to the star's LSR (``apex velocity'') ${\mathbf v}_\odot$, 
whose absolute value is listed in Table~\ref{tab:enc} for each stellar category. We assume 
a random relative orientation of the two vectors and thus compute:
\begin{equation}\label{eq:velocity}
V=\bigl\{v_{\odot i}^2+v_\ast^2-2v_{\odot i}v_\ast\cdot C\bigr\}^{1/2}
\end{equation}
where $C=\cos\theta$ is taken as $C=2\xi_C-1$, and $\theta$ is the angle between the two
vectors.

Within each stellar category we have to account for the fact that the contribution to the 
encounter flux is proportional to $V$. Thus we define a constant, large velocity  $V_{0i} = v_{\odot i} + 3\sigma_{\ast i}$
for each category, such that $V$ is always smaller than $V_{0i}$, and we take a new random 
number $\xi_V$ and keep the value just found for $V$, if $\xi_V<V/V_{0i}$. Otherwise we 
repeat the computation of $V$ until the $\xi_V$ condition is fulfilled. This procedure 
was not followed in our previous investigations, leading to underestimates of the average 
stellar velocities. Further underestimates were caused by programming errors, and we 
caution the reader that the stellar velocities in \citeauthor{RICetal:04}~\shortcite{RICetal:04,RICetal:05}
were systematically too small. This is clearly seen by comparing Fig.~1 of \inlinecite{RICetal:05} with 
the data in Table~\ref{tab:enc}, which yield a mean velocity of 53~km/s with a dispersion of 
$\simeq20$~km/s.

A few comments on the mean stellar encounter velocity are in order. The average peculiar 
velocity of stars in the solar neighbourhood is $\simeq40$~km/s. This value was given by 
\inlinecite{HUT.TRE:85}, and combining it in quadrature with a typical solar apex velocity 
of 23~km/s for the most common stellar categories (Table~\ref{tab:enc}), one gets a mean 
heliocentric velocity of $\simeq46$~km/s. \inlinecite{HEIetal:87} were the first to 
introduce the flux-weighting into the selection of random velocities, but they neglected 
the solar apex velocity. In fact, their flux-weighting was somewhat different from ours, 
because they considered only one of the three velocity components, namely, the radial
heliocentric velocity. But the encounter flux is sensitive to the velocity with respect to 
the impact plane, \ie, the full speed of the star, instead of just the radial component. We 
have found that this difference has only a small effect on the resulting mean velocity, but 
we mention it for the sake of completeness. In both cases, we find that the weighting 
raises the mean velocity by $6-7$~km/s. This explains our mean velocity of 53~km/s as
resulting from including both the solar motion and the flux-weighting procedure. Finally, 
let us compare with the mean encounter velocity of $\simeq46$~km/s in \inlinecite{GARetal:01}.
This resulted from a list of 92 stellar encounters within 5~pc and 1~Myr of the present,
compiled with the aid of Hipparcos data, but the authors showed that there was a serious 
bias against faint absolute magnitudes in this sample, affecting all stars with $M_V>4$.
Thus, the stars with the highest velocities were essentially lacking, and the resulting mean 
velocity is that of the inherently brighter, slower moving stars.

Comparing with other investigators, we note that both \inlinecite{HEI:90}, 
\inlinecite{MAZ:06} and \inlinecite{EMEetal:07} based their stellar encounter frequencies 
on the analysis by \inlinecite{HEIetal:87}, who ignored the solar motion -- thereby 
underestimating the relative frequency of encounters with massive stars that have small 
velocity dispersions -- and neglected the contribution of the massive giants. Our encounter 
sample contains as much as 3.5\% of massive stars, \ie, the B0, A0, A5, F0, F5 
and `gi' categories in Table~\ref{tab:enc} with an average mass of 2.3~$M_\odot$, while 
counting the stars in the absolute magnitude range that corresponds to this average mass 
in \inlinecite{HEIetal:87}, one arrives at $<1$\% of the total encounter frequency. Since the 
massive stars have an average $M/V$ ratio $\sim 10$ times larger 
than the red dwarf stars that dominate the sample, each such star will affect $\sim 100$ 
times as many comets. Hence one easily realizes that in our case a large fraction of the 
total stellar effect comes from the massive star category that is downplayed by 
the other investigators. This, to some extent, compensates for two other effects that make 
the stellar perturbations less efficient in our simulation. One is the larger encounter 
velocities, as already described, and the other is the total encounter frequency within 
1~pc, which in our case is 10.5 per Myr, while for the others it is 13.1 per Myr.

Our calculations of the heliocentric impulse imparted to the comet (at time $t$) are done 
using the Sequential Impulse Approximation, which guarantees a good accuracy at a low 
cost of computing time \cite{RICetal:05}.
During the simulation we keep track of all the perihelion passages with their $q$ values.
At each perihelion a decision is taken about which method to use for the Galactic tide
perturbation during the coming orbital period. The fastest method is a mapping \cite{BREetal:07}
that analytically computes the orbital elements at the subsequent perihelion,
but this is used only for elliptic orbits with semi-major axis less than a critical value
that depends on the eccentricity \cite{FOUetal:07b}. Otherwise we use numerical
integration with a symplectic integrator for KS-regularized equations of motion \cite{LAS.ROB:01}
in case $1/a>10^{-5}$~AU$^{-1}$ and the 15$^{th}$ order RADAU
integrator \cite{EVE:85} for $1/a<10^{-5}$~AU$^{-1}$.

During the orbital period in question, normally, several stellar passages occur. On
those occasions the osculating cometary orbit is subject to an instantaneous impulse.
In the numerical integration regime for the Galactic tide, one always comes back to
perihelion. But in the mapping regime, when the starting orbital period has elapsed, 
the comet may not be at perihelion because of the intervening stellar perturbations, and
we then resort to numerical integration until the next perihelion passage takes place.

The simulation proceeds for a maximum of 5~Gyr, unless an end state is reached.
There are two such end states: either the comet reaches perihelion with $q<q_c=15$~AU 
(it is lost due to planetary perturbations), or the comet reaches
$r=4\times 10^5$~AU in outbound motion (it escapes directly into interstellar space).

What we have just described is the full simulation of the ``combined'' model 
including both the Galactic tide and stellar perturbations. In addition, we have run two 
simulations that include only one or the other of the two dynamical effects.

\section{Injection time}

We will first consider the time needed to shift any cometary starting perihelion distance 
$q_o\,>\,32$ AU into a perihelion distance $q\,<\,15$ AU. This is the time $t_{\rm inj}$ 
required to inject a comet into the target zone, and we call it the injection time. We 
have thus scrutinized all three simulations, and for each injected comet in every 
simulation we noted its value of $t_{\rm inj}$. Let us now compare the statistics of 
injection times between all three dynamical models.

The range of initial semi-major axes $3\,000\,<\,a_o\,<\, 100\,000$~AU is divided, 
according to $\log a_o$, into ten equal intervals. For each interval the following 
statistical parameters concerning the injection time are computed: its median 
value, its lower and upper quartiles (surpassed by 75\%, respectively 25\%, of the 
values), its lower and upper deciles (surpassed by 90\%, respectively 10\%, of the 
values), and finally its lower and upper percentiles (surpassed by 99\%, respectively 
1\%, of the values).

\begin{figure}[ht!]
\begin{center}
\setlength{\unitlength}{1.cm}
\begin{picture}(10,12)
\put(0,6){\includegraphics[height=6cm]{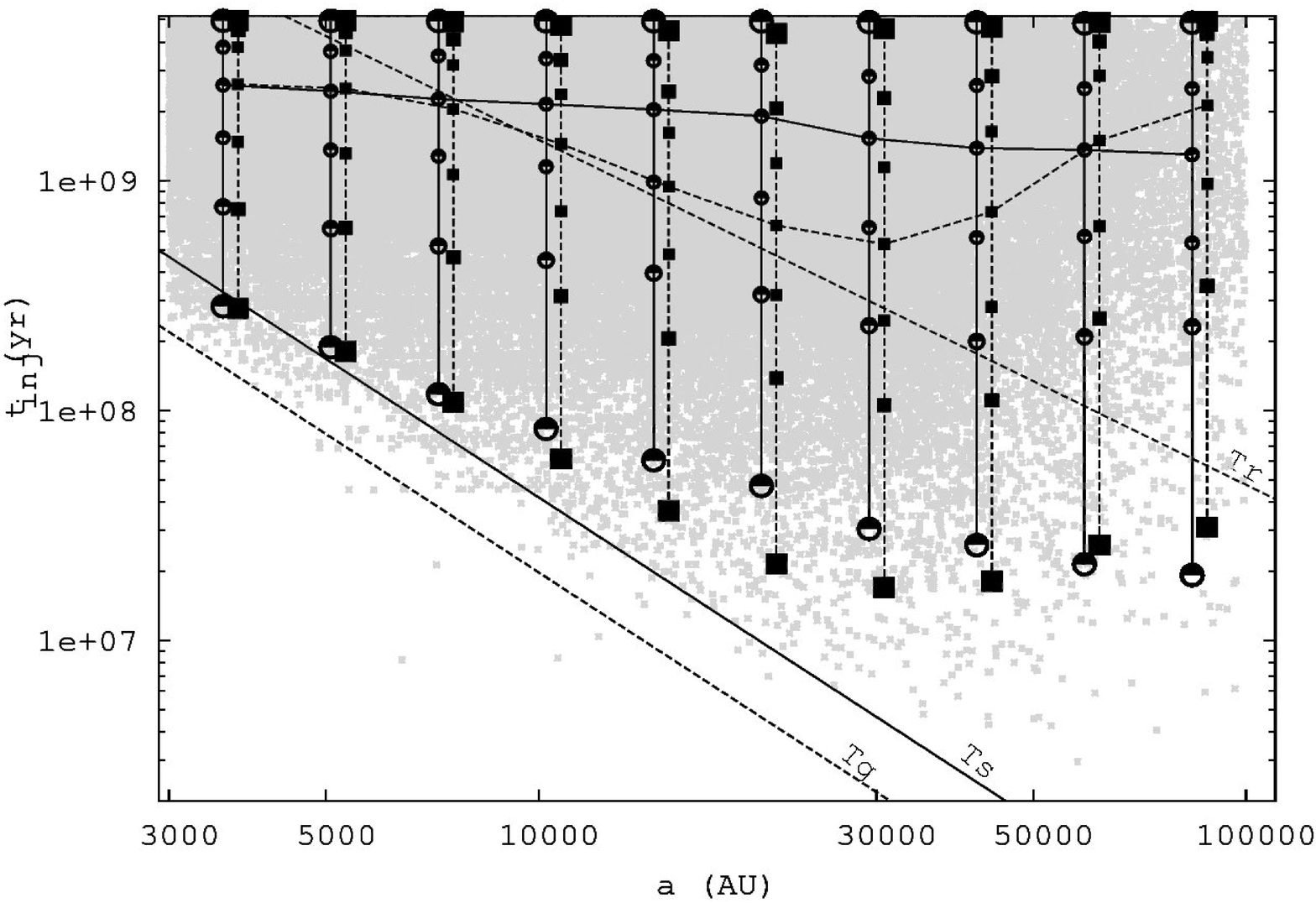}}
\put(0,0){\includegraphics[height=6cm]{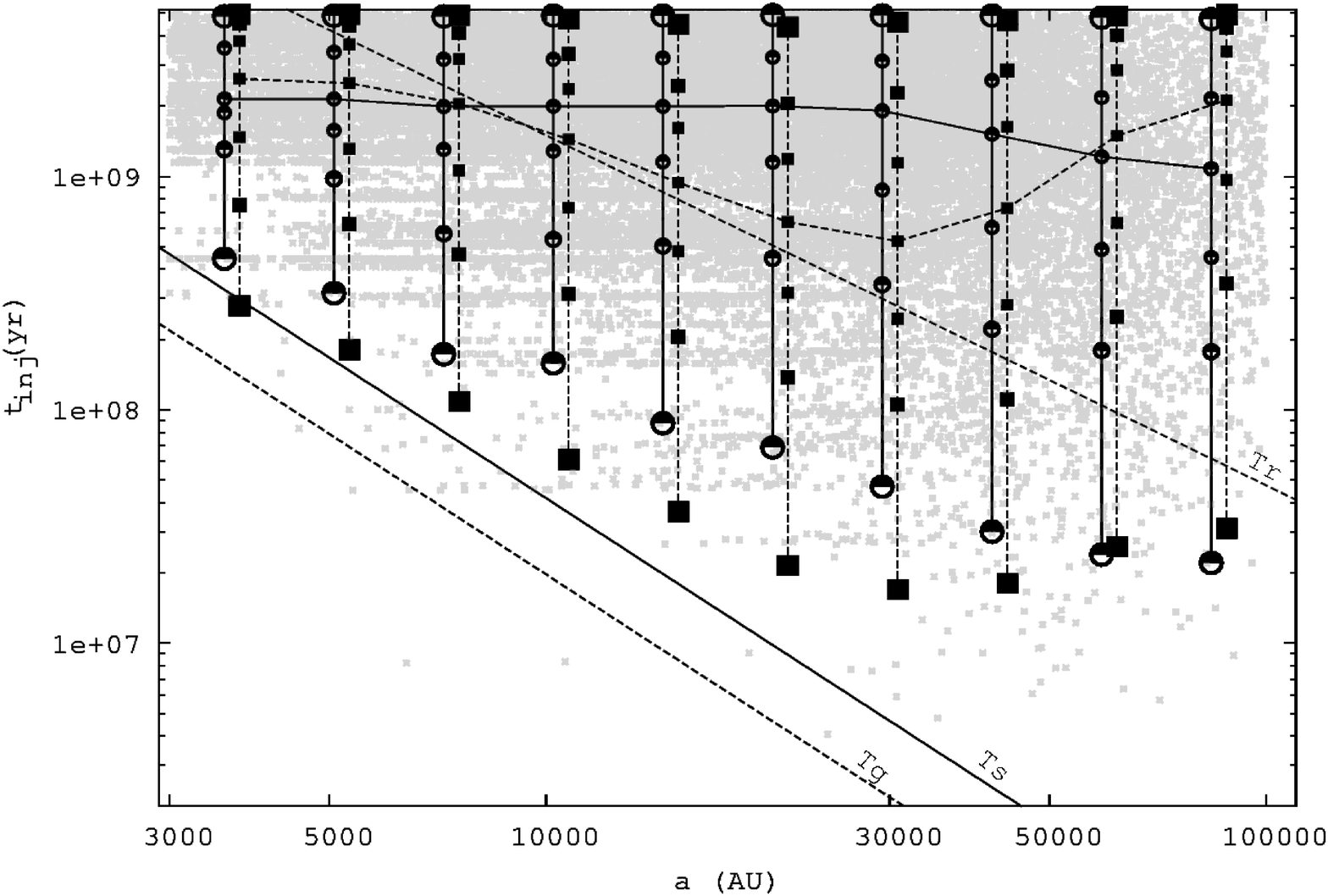}}
\end{picture}
\caption{Injection time versus initial semi-major axis. The semi-major axis range 
is split into ten equal intervals of $\log a$. For each interval we plot different 
statistical parameters characterizing the distribution of injection times, as 
explained in the text. The dotted line labelled $T_g$ corresponds to the tidal 
torquing time shown in Fig.~2 of Duncan {\it et al.} (1987), $T_s$ to the 
corresponding stellar torquing time, and $T_r$ to the period of ($q,i_G$) oscillation 
imposed by the vertical tide. The left panel (Fig.~\ref{fig:dqt}a) compares the model 
with only tides to the one with only stars, and the grey dots are individual stellar 
injections. In the right panel (Fig.~\ref{fig:dqt}b) the stars-only model is replaced 
by the combined model.}\label{fig:dqt}
\end{center}
\end{figure}

Figure~\ref{fig:dqt} presents the comparison of $t_{\rm inj}$ statistics by means of two plots. 
The left one (Fig.~\ref{fig:dqt}a) compares the model with only the Galactic tides to 
the one with only stellar perturbations, while in the right one (Fig.~\ref{fig:dqt}b) 
the tides-only model is compared to the combined model. In each case we plot the 
statistical quantities versus $a_o$. The tides-only model is represented by filled 
squares, and for the other models we use half-filled circles. At any particular 
value of $a_o$, the symbols for each model are joined by vertical bars. A slight
horizontal shift between the symbols has been introduced for easy distinction of the 
models, but the real $a_o$ intervals are identical. The median values have been joined 
by curves (dotted for the tides-only model, and solid for the other ones). The grey 
dots show individual injections for the stars-only model (Fig.~\ref{fig:dqt}a) and 
the combined model (Fig.~\ref{fig:dqt}b).

For comparison with~\inlinecite{DUNetal:87}, who plotted a similar diagram (their 
Fig.~2), we include three lines. The one labelled $T_g$ shows their ``tidal torquing 
time'' and the one labelled $T_s$ shows the corresponding ``stellar torquing time'', 
both as functions of $a$. These are meant to represent the typical time 
required to decrease the perihelion distance from 25~AU to 15~AU in the two cases. 
The third line labelled $T_r$ refers to the period of ($q,i_G$) oscillation due to 
the tidal component normal to the Galactic plane \cite{DUNetal:87}.

By inspecting Fig.~\ref{fig:dqt}, we can make the following observations. First, 
compare the tides-only median curve with the $T_r$ line. The two agree fairly well in 
the range from $a\simeq6000$ to $25\,000$~AU. This is natural, because $T_r$ is twice 
the average time it takes for the vertical tide to bring any Oort Cloud comet into the 
target zone, as long as it is on a relevant trajectory with $q_{\rm min}<15$~AU. For 
$a\lta6000$~AU the median flattens out at about 2.5~Gyr, and this is due to the limit 
of our simulated interval at 5~Gyr. Had we let the simulation run for a much longer 
time, we would have seen the median curve follow the $T_r$ line to even smaller
$a$ values. For $a\gta25\,000$~AU we see how the median curve turns upwards, while 
$T_r$ continues to decrease. This can be explained as a result of a quick stripping 
of comets from all trajectories with $q_{\rm min}<15$~AU, after which these have to 
be repopulated through the action of the non-integrable part of the tides. Since this 
works on a much longer time scale, it is obvious that the median of $t_{\rm inj}$ has 
to increase.

Already at this point we see evidence that the mean injection time -- even in the 
tides-only model -- does not follow the prediction of the $T_r(a)$ dependence at all 
semi-major axes. Studying the Oort Cloud over a long enough time allows other parts 
of the Galactic tide than the simple, vertical component to take control of comet 
injections, at least in the outer parts of the cloud. But consider now the 
median curves of the two models that involve stellar perturbations. They are mutually 
quite similar, but they differ strongly from the tides-only curve except at $a\lta 
6000$~AU.

The mutual similarity -- in spite of a much larger number of grey points (injections) 
in Fig.~\ref{fig:dqt}b -- means that the same basic mechanism is at work. We identify 
this as the angular momentum reshuffling by stellar perturbations. In Fig.~\ref{fig:dqt}a 
(stars-only model) this in itself makes comets diffuse all over angular momentum space 
so that some reach the target zone. In Fig.~\ref{fig:dqt}b (combined model) the same 
angular momentum diffusion repopulates the ``infeed trajectories'' (with $q_{\rm min} 
<15$~AU) of the vertical Galactic tide, whereupon the comets are injected at a rate given 
by $T_r(a)$. We interpret the flatness of the median curves at a level of roughly half 
the duration of our simulation as evidence that the time scale of angular momentum 
reshuffling is short enough to guarantee an injection rate that is at least as large
during the first half ($0-2.5$~Gyr) as during the second half ($2.5-5$~Gyr). This is 
in agreement with the thermalization time scale reported by \inlinecite{DUNetal:87}. 
The tendency for a slight decrease of the median $t_{\rm inj}$ at the largest 
semi-major axes is likely due to a progressive depletion of the outer parts of the cloud 
during the course of the simulation, while the reshuffling time scale is relatively 
short.

We thus realize that the behaviour of the median injection time, generally speaking, has 
very little to do with any of the theoretical time scales. Let us now instead consider the 
lowest percentiles, since these give information on the quickest injections that -- in 
principle -- might be governed by the $T_g$ or $T_s$ dependences. The lowest percentile 
of the tides-only model indeed decreases with $a_o$ roughly parallel to the $T_g$ line 
in the inner core of the cloud ($a<10\,000$~AU), but we nonetheless see a somewhat 
smaller slope. This tendency gets stronger with increasing $a_o$ and finally turns into 
an increase outside $30\,000$~AU. We interpret this as due to the same repopulation of 
infeed trajectories by the non-integrable part of the tide as we discussed in relation 
to the median curve.

The lowest percentile of the stars-only model shows a fall-off with $a_o$ that is 
interestingly slow in comparison with the $T_s$ line. This appears to be related to the 
horizontal bands of grey points, which are cometary showers. For each $a_o$ interval in 
the inner core, the timing of the lowest percentile is that of the first shower reaching 
into that interval. The larger $a_o$, the sooner such a shower appears. But the showers 
also get weaker, being caused by more and more distant stellar encounters. Thus, in the 
outer parts of the cloud they are no longer of significance for defining the lowest 
percentile. Since this is instead controlled by a growing number of usual, inefficient 
stars passing through the outer regions, one has to wait longer.

When we look at the lowest percentile of the combined model, we see that it follows the 
same decrease as the tides-only model in the inner core. Indeed, with a much larger 
number of injections, the comet showers have lost their importance, and as we shall see 
in Sect.~4, during the first Gyr the injections are largely controlled by the Galactic 
tides. But outside the inner core the lowest percentile now behaves with respect to that 
of the tides-only model in a similar way as the median does, and the reason is the same. 
Going to larger semi-major axes, in both models we see an increasing number of late 
injections, although for different reasons, and these determine the behaviour of most 
statistical parameters, causing them to decrease less rapidly than the $T_g$ line.

In summary we can state that we have found the theoretical time scales of 
\inlinecite{DUNetal:87} to give a rather poor representation of our statistics of 
injection times. On the other hand, we find evidence in the combined model for an 
important role being played by the repopulation of tidal infeed trajectories via 
stellar encounters -- something that may be described as a synergy effect. This being 
said, one nonetheless realizes that $T_r$ is one of the basic time scales that govern 
this synergetic injection, the other one being the angular momentum reshuffling time 
scale of stellar perturbations. Let us now move to a discussion of the rate of 
injections and how this depends on time.

\section{Time dependent injection flux}

\begin{figure}[ht!]
\begin{center}
\psfrag{tau}{\small $\tau (\%)$}
\psfrag{NC-NS-NG}{\small $\Delta N_C$}
\psfrag{NOC}{\small $N_{OC}\times 10^{-5}$}
\psfrag{N}{\small $N$}
\psfrag{t}{\small $t$~(Gyr)}
\includegraphics[width=12.cm]{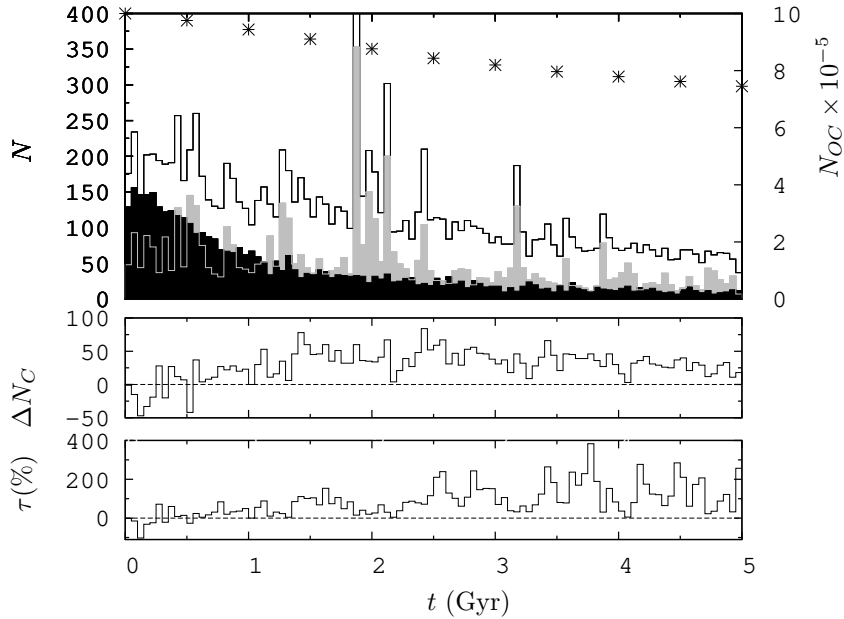}
\caption{The upper diagram shows the number of comets entering the observable zone 
per 50~Myr versus time. The white histogram corresponds to the combined model, the black 
histogram to the Galactic tide alone, and the grey histogram to the passing stars alone. 
The asterisks indicate the number of comets remaining in our simulation 
for the combined model at every 500~Myr 
with scale bars to the right. The middle diagram shows the excess number of injections 
into the observable region per 50~Myr in the combined model with respect to the sum of 
the stars-only and tides-only models. The lower diagram shows this excess expressed in 
percent of the mentioned sum.
 }\label{fig:flux_all}
\end{center}
\end{figure}

The upper part of Fig.~\ref{fig:flux_all} shows a histogram plot of the number 
of comets injected into the observable region as a function of time 
from the beginning till the end of the simulation. Three histograms are shown 
together: the one in black corresponds to a model with only Galactic tides, and 
the grey one to a model including only stellar perturbations. Finally, 
the top, white histogram is for the combined model that includes both tides and 
stars.

The first thing to note is the gradual decline of the injection flux of the tides-only model
over a few Gyr, after which it stays at a very low level. The reason is clear. In the
beginning, the phase space trajectories that in the regular dynamics imposed by the
vertical tide will periodically lead into the ``loss cone'' ($q<15$~AU) are populated just as
densely as any phase space domain and thus furnish an important flux of injections
during the first period of ($e,i_G$) oscillations. This amounts to a typical time scale of $<
1$~Gyr for much of the initial cloud, but the population on these trajectories is depleted
by each entry into the loss cone, and there is no efficient way to replenish them
without including stellar passages. On the longer time scale, we see only the feeble
flux coming from (1) the infeed into the tidal injection trajectories by the 
non-integrable part of the tide; (2) the inner parts of the cloud, where the period of 
oscillation is very long \cite{FOUetal:06}.

The other two histograms include the effects of stellar passages, and the stars are
the same in both simulations. Therefore, we see the same comet showers appearing
and the same quasi-quiescent periods in between. The white area at the top of each bin 
corresponds to the extra contribution of the combined model as compared with that of the 
stars only. If the numbers plotted in the white, grey and black histograms are called 
$N_C$, $N_S$ and $N_G$, respectively, we can define $\Delta N_C = N_C - N_S - N_G$ as an 
absolute measure of this extra contribution.\footnote{Towards the end of our simulation 
the number of Oort Cloud comets has decreased in all three models but most in the combined 
one. We then have about $930\,000$, $840\,000$ and $760\,000$ comets in the tides-only, 
stars-only and combined models, respectively. This means that $\Delta N_C$ actually underestimates 
the extra contribution of the combined model.} Already at first glance, looking at the later 
part of the simulation, we see that this is very significant. In the two lower panels of 
the Figure, we plot histograms of $\Delta N_C$ and $\tau = \Delta N_C/(N_S+N_G)$, \ie, 
the extra contribution expressed as a fraction of $N_S+N_G$.

\begin{figure}[ht!]
\begin{center}
\psfrag{NC-NS-NG}{\small $\Delta N_C$}
\psfrag{NS}{\small $N_S$}
\includegraphics[width=8.cm]{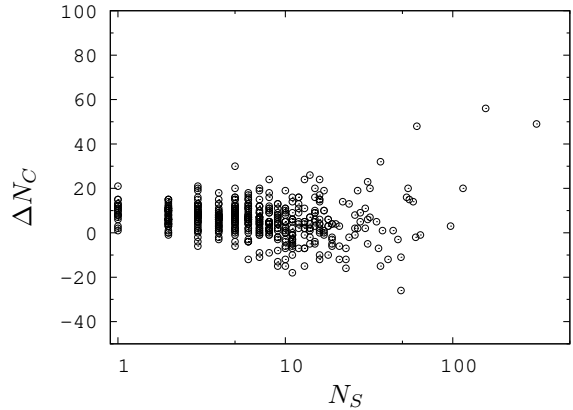}
\caption{$\Delta N_C$ versus $N_S$ (plotted on a log scale), where the numbers refer to 
injections of comets 
into the observable region during intervals of 10~Myr.}\label{fig:cor_syn1}
\end{center}
\end{figure}

The basic observations are as follows. The spiky nature of the grey histogram is due to 
comet showers caused by close stellar encounters (we will briefly treat these below). 
While during the first Gyr the level of $N_G$ is generally higher than that of $N_S$, 
this situation gets reversed after more than two Gyr. Even outside the main showers, 
$N_S$ is then at a somewhat higher level than $N_G$. The white histogram, showing $N_C$, 
shares the spikes of the strongest showers, but the contrast between the spikes and the 
background is less than in the grey histogram. Indeed, the $\Delta N_C$ histogram shows 
no spikes at all, and the general level does not seem to correlate with the stellar 
injection rate, as illustrated by Fig.~\ref{fig:cor_syn1}. Therefore, during the later 
part of the simulation, the $\tau$ parameter shows fluctuations anticorrelated with those 
of $N_S$. It reaches a few hundred percent, when $N_S$ drops to its lowest levels, but 
sometimes decreases to nearly zero during the peaks of $N_S$.

In order to smooth out those fluctuations we present in Table~\ref{tab:tau} time averages 
of $\tau$ over 1~Gyr periods along with the corresponding integrals of $N_C$, $N_S$ and 
$N_G$. During the first Gyr the flux of the combined model is not much larger than the 
sum of the two fluxes with separate effects, and the difference is just a small fraction 
of the total flux. But toward the end the synergy effect of the combined model, as 
measured by $\Delta N_C$, has grown -- on the average -- to nearly the same level as 
$N_S+N_G$. During the last Gyr we find that $\langle N_C\rangle$ is about 2.5 times 
larger than $\langle N_S\rangle$ in fair agreement with earlier estimates by 
\inlinecite{HEIetal:87} and \inlinecite{HEI:90}. After an initial, relatively fast 
decrease due to the emptying of the tidal infeed trajectories, $\langle N_C\rangle$ 
continues to decrease approximately in proportion to the total number of Oort Cloud 
comets ($N_{OC}$), and $\langle N_S\rangle$ and $\langle N_G\rangle$ show similar 
behaviours.

\begin{table}
\begin{tabular}{|c|ccccc|}
Model & $ [0 - 1]$~Gyr & $ [1 - 2]$~Gyr & $ [2 - 3]$~Gyr & $ [3 - 4]$~Gyr & $ [4 - 5]$~Gyr \\
\hline 
G     & 2\,128     & 797     & 481   & 307    & 248    \\
S     & 1\,425     & 1\,555 & 1\,030 & 717    & 511    \\
C   & 3\,618     & 3\,141 & 2\,412 & 1\,733 & 1\,274 \\
$\langle\tau\rangle$ & 1.8\%    & 33.6\% & 59.6\% & 69.2\% & 67.9\% \\
\hline
\end{tabular}
\caption{Number of comets entering the observable region during periods of 1~Gyr.
Model G corresponds to the Galactic tide alone, S to passing stars alone, and C to
Galactic tide and passing stars together. $\langle\tau\rangle$ is the increment 
from the sum of the two first rows (Galactic tide plus passing stars separately) to 
the third row (Galactic tide and passing stars together).
}\label{tab:tau}
\end{table}

Looking in detail at the $\Delta N_C$ and $\tau$ histograms in Fig.~\ref{fig:flux_all} 
for the beginning of our simulation, we see that they start from negative values and 
turn into positive ones after $\sim0.5$~Gyr. Thus, in the very beginning, the sum of 
the separate fluxes is larger than the combined flux. This phenomenon was found by 
\inlinecite{MAT.LIS:02}, whose calculations were limited to only 5~Myr, and as they 
explained, it is typical of a situation where both tides and stars individually are able 
to fill the loss cone to a high degree. We will discuss this point again in Sect.~5.

The large amount of synergy ($\tau\sim70$\%) seen in the later part of our simulation 
is remarkable and indicates that both the tides and the stars on their own are seriously 
inefficient in filling the loss cone. It is only by means of the synergy of both effects 
that we are able to explain an important degree of loss cone filling at the current 
epoch. We will look at this closer in Sect.~5 by separating the injection flux into 
different ranges of semi-major axis. For the moment we emphasize that {\it treating comet 
injections from the Oort Cloud in the contemporary Solar System simply as a result of the 
Galactic tide is not a viable idea}.

Already in Sect.~3 we identified a mechanism that offers a likely explanation of the 
synergy effect, \ie, the repopulation of tidal infeed trajectories via stellar encounters. 
But note in Fig.~\ref{fig:flux_all} that the initial flux of the model with tides only is 
not matched by the white areas in the later part of the simulation. Thus, even though 
there is an ongoing replenishment of the tidal infeed trajectories due to the randomizing 
effect of stellar encounters, this replenishment is not complete. {\it The critical 
trajectories remain largely depleted, and models that do not take this fact into account 
will overestimate the tidal contribution to the injection flux, as well as the efficiency 
of tides in filling the loss cone.}

The most important synergy mechanism of the Galactic tide and stellar perturbations is
that the latter are able to repopulate the critical phase space trajectories that in
the quasi-regular dynamics imposed by the tide lead into the loss cone \cite{DYB:02,FER:05}. 
Our results appear to verify and quantify this picture. But in addition we see hints that 
a different effect is also at work. In Sect.~5 we will show that the distribution of 
$1/a$ of newly injected, observable comets -- even during the typical, quiescent periods -- 
has a significant extension inside the limit (at $a\simeq3\times10^4$~AU), where the tide 
becomes able to feed comets from outside the loss cone into observable orbits. Our 
explanation for this effect is as follows.

In qualitative terms, when the Galactic tide is in the process of injecting a comet into 
the observable region from the region outside the loss cone, and stellar perturbations are 
added, the latter will sometimes aid in decreasing the perihelion distance of the comet 
($\Delta q_\ast < 0$), and sometimes they will counteract the tide ($\Delta q_\ast > 0$). 
To first order, the two effects will cancel. But if we consider how much the critical value 
($1/a_c$) of the inverse semi-major axis for tidal injection into the observable region 
($a_c\simeq 3\times10^4$~AU) is decreased by a typical positive $\Delta q_\ast$ or increased 
by the same typical value of $\Delta q_\ast$ in the negative direction, we find that the 
latter effect dominates. Thus, a negative $\Delta q_\ast$ causes a larger gain of comets 
with $a<a_c$ than the loss of comets with $a>a_c$ caused by a positive $\Delta q_\ast$ of 
the same size. This holds for any nearly uniform distribution of $1/a$ in the Oort Cloud.

In mathematical terms, consider the expression for the maximum possible decrease of $q$ 
in one revolution due to the Galactic tide:
\begin{equation}\label{eq:max_pert}
\Delta q = - S q^{1/2} z^{-7/2}
\end{equation}
where $z=1/a$ and $S=2.8\times10^{-15}$ \cite{BYL:86}, counting $q$ and $a$ in AU. This 
would hold for a Galactic latitude of perihelion of $\pm45^\circ$. We take this 
favourable orbital orientation as an example, but the following arguments apply for any 
other orientation as well.

Next, consider a particular value ($q_p$) of the perihelion distance preceding the injection 
into an observable orbit. Using Eq.~(\ref{eq:max_pert}), one can write down an approximate 
condition for the critical value $z=z_0$ in order to bring the comet into the observable 
region ($q<q_\ell=5$~AU):
\begin{equation}\label{eq:ener_lim}
q_p = q_\ell + S q_p^{1/2} z_0^{-7/2}
\end{equation}
Eq.~(\ref{eq:ener_lim}) defines a unique relation between $q_p$ and $z_0$, and by 
differentiating one easily finds that $q_p$ decreases monotonously with $z_0$, while the 
second derivative is always positive.

Considering thus an arbitrary point ($z_0,q_p$) satisfying Eq.~(\ref{eq:ener_lim}), we may 
introduce stellar perturbations by adding a term $-\Delta q_\ast$ to the right-hand member 
of Eq.~(\ref{eq:ener_lim}), and we can write:
\begin{equation}\label{eq:el_stars}
q_p = q_\ell + S q_p^{1/2} z_1^{-7/2} - \Delta q_\ast
\end{equation}
where $z_1$ is the new critical value of $z$. Hence, ($z_1,q_p+\Delta q_\ast$) also
satisfies Eq.~(\ref{eq:ener_lim}). Without the stellar perturbation all comets with 
$q=q_p$ and $z<z_0$ are injected into $q<q_\ell$, and including the stellar perturbation,
the condition changes into $z<z_1$. From the negative slope of the $q_p(z_0)$ relation it 
is obvious that a negative value of $\Delta q_\ast$ yields $z_{1+}>z_0$, and the same 
positive value yields $z_{1-}<z_0$. It is also obvious from the positive curvature of the 
graph that $z_{1+}-z_0>z_0-z_{1-}$. If $\vert\Delta q_\ast\vert$ is very small, this 
difference is negligible, but if it is large enough to compete with $\vert\Delta q_G\vert$, 
the effect will be important. The latter is indeed often the case, when we discuss 
injections from just outside the loss cone ($q_p\simeq15$~AU), as has been shown, \eg, by
\inlinecite{DUNetal:87} and \inlinecite{FER:05}. Therefore, the gain of comets with $z>z_0$ 
occurs over a larger interval than the loss with $z<z_0$ for a symmetric distribution of 
stellar perturbations.

Another issue is the 
distribution function of $z$ for the Oort Cloud. Our simulations start with a probability 
density $f(z)\propto z^{-1/2}$ as appropriate for an Oort Cloud formed according to the 
model of \inlinecite{DUNetal:87}. In such a situation there would be more comets per 
unit interval of $z$ at $z<z_0$ than at $z>z_0$, and the gain effect would be 
counteracted and possibly turned into a net loss of injected comets.

However, an interesting result of our simulations is that the gradual loss of comets 
from the Oort Cloud changes the distribution of $1/a$. In agreement with a recent study by \inlinecite{BRAetal:08}, we find
that the loss of comets from the outer 
parts of the cloud is not compensated by diffusion from the inner parts, so that after 
4.5~Gyr, when the number of comets has decreased from $1\times10^6$ to $7.6\times10^5$, 
$f(z)$ has become roughly flat over the range from $a=20\,000$ to $100\,000$~AU. This 
shows that we have to expect a net gain of newly injected, observable comets resulting 
from a synergy of $\Delta q_G$ and $\Delta q_\ast$. Moreover, there should be a shift 
of comets from outside to inside the tidal injection limit -- probably explaining why 
we see a significant flux of new comets all the way down to $a\simeq20\,000$~AU.

Although we cannot provide exact numbers, it appears that the secondary synergy 
mechanism due to what we may call ``constructive interference'' of the two effects -- 
even though it certainly exists -- is not the dominating one. The tentative evidence 
comes from the relative smoothness of the $\Delta N_C$ histogram (Fig.~\ref{fig:flux_all}) 
and the lack of correlation between $\Delta N_C$ and $N_S$ (Fig.~\ref{fig:cor_syn1}). 
These properties 
are expected of the repopulation of tidal infeed trajectories because of the long 
response time ($\sim$ several $10^8$~yr) for tidal infeed on the $T_r$ time scale 
(Fig.~\ref{fig:dqt}). But if the constructive interference had been very important, we 
would have expected $\Delta N_C$ to increase immediately upon an increase of $N_S$ --
with the caveat that visible peaks of our $N_S$ histogram might arise primarily from 
a temporary infeed of inner Oort Cloud comets with $z>>z_o$, which do not contribute 
to the interference. Further detailed studies are needed to clarify this issue.

\begin{figure}[ht!]
\begin{center}
\psfrag{N}{\small $N_C$}
\psfrag{t}{\small $t$~(Gyr)}
\includegraphics[width=12.cm]{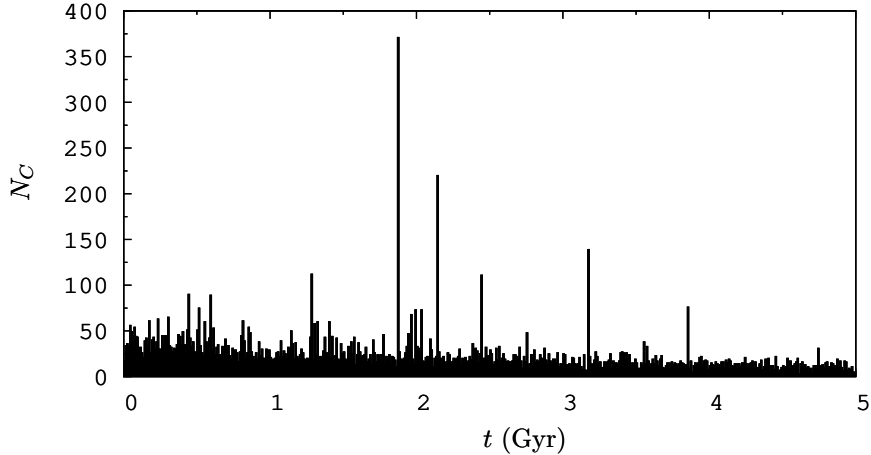}
\caption{Same as Fig.~\ref{fig:flux_all} (upper panel), but versus time per 10~Myr for 
the combined model only.}\label{fig:flux_10d6}
\end{center}
\end{figure}

The cometary showers, displayed in Fig.~\ref{fig:flux_10d6} by means of a histogram 
of the injection flux of the combined model with a bin width of only 10~Myr, are seen 
to be quite important for the injection of comets from the Oort Cloud, as expected and 
as found by other authors previously (\eg, \opencite{HEIetal:87}). We are saving a 
detailed analysis of those for a later paper. At present, we can only remark that the 
results presented here are hard to compare with the treatments of cometary showers by 
\inlinecite{HEIetal:87} or \inlinecite{HEI:90}. The first of these papers treated only 
comets with $a=10\,000$ or $20\,000$~AU with a full dynamical model and then only for a 
time interval of less than 200~Myr. The second gave only a brief account of a simulation 
for 4.5~Gyr and then only for injections from $q>10$ to $q<10$~AU instead of our 
requirement that the perihelion has to fall substantially from $q>15$ to $q<5$~AU.

\section{Orbital element distributions of observable comets}

\begin{figure}[ht!]
\begin{center}
\setlength{\unitlength}{1.cm}
\psfrag{-1/a}[cc]{\tiny $-1/a\times 10^5~{\rm (AU^{-1})}$}
\psfrag{N}{\tiny $$}
\psfrag{sinb}{\tiny $|\sin b|$}
%\psfrag{-1}{}
\begin{picture}(11.8,6.)
\put(7.54,0.){\includegraphics[width=4.cm,height=6.cm.]{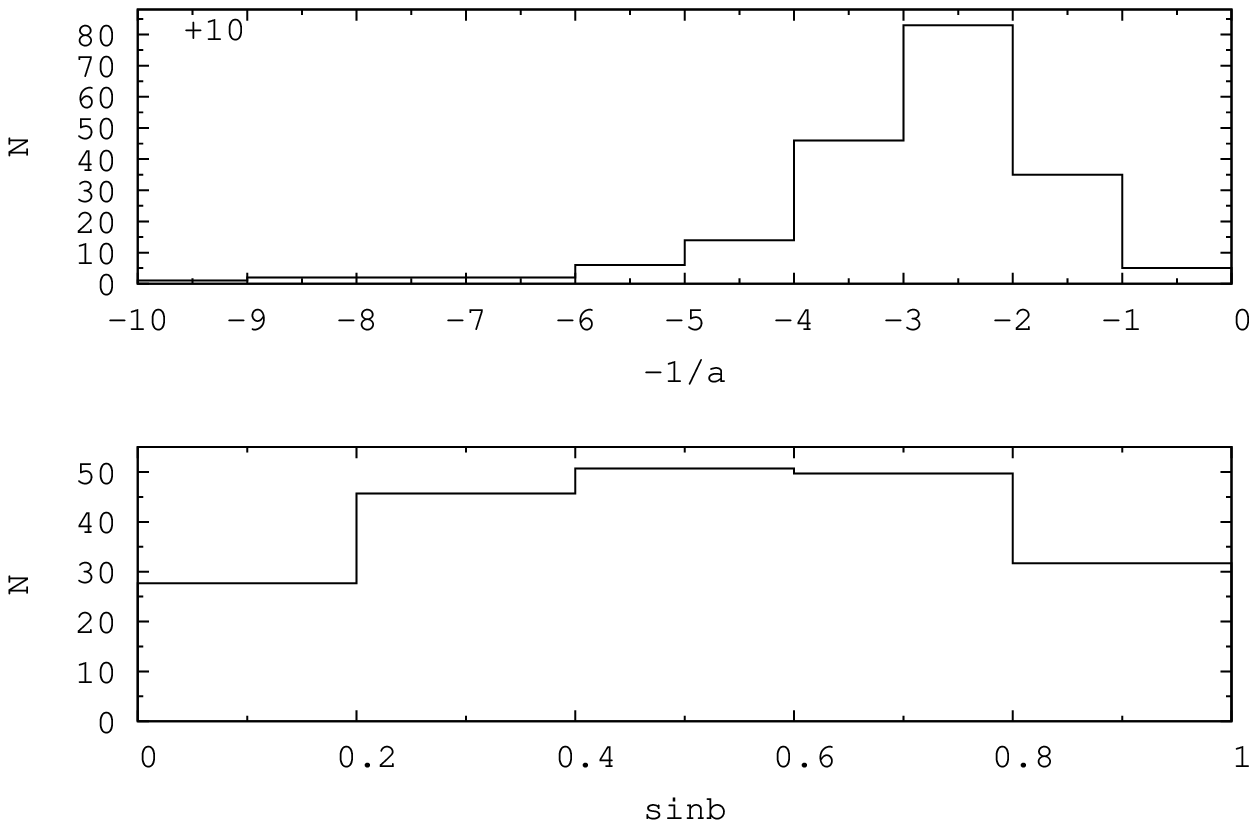}}
\put(3.85,0.){\includegraphics[width=4.cm,height=6.cm]{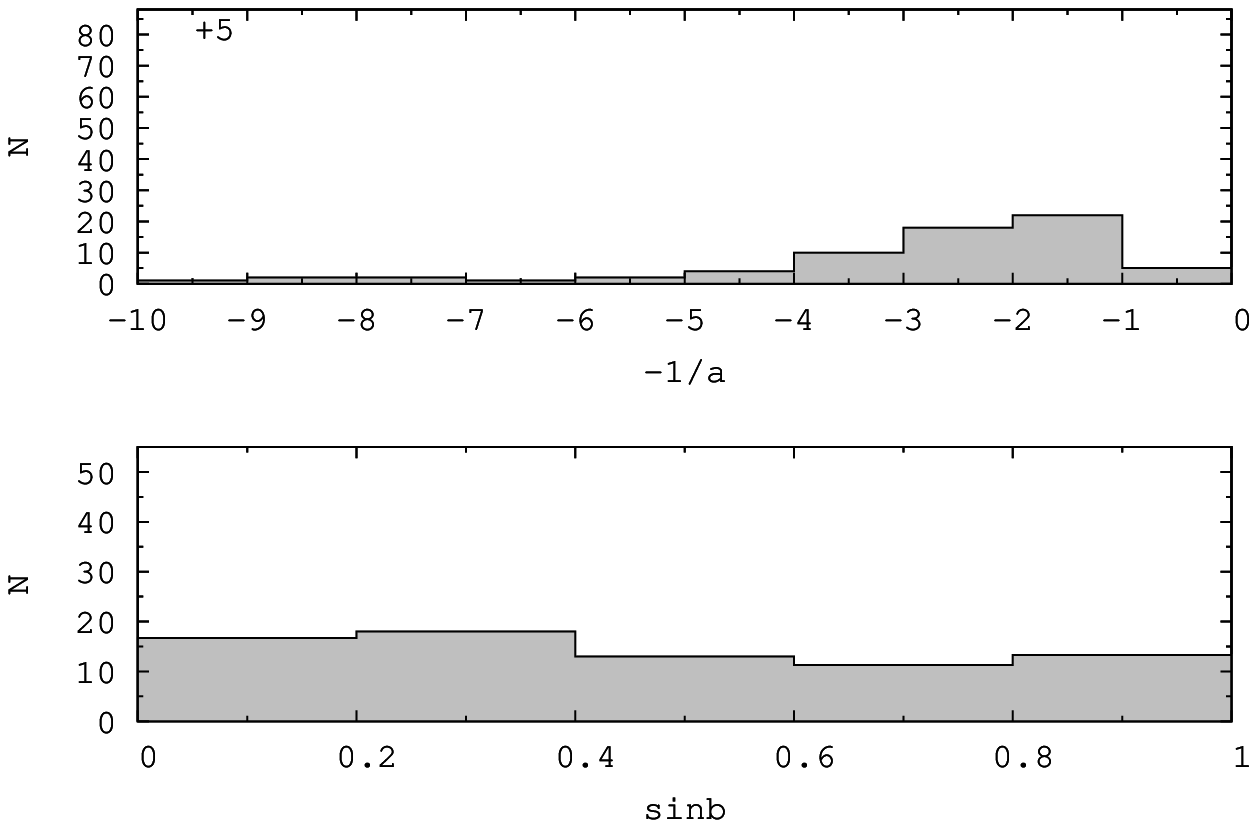}}
\psfrag{N}{\tiny $N$}
\put(0,0.){\includegraphics[width=4.cm,height=6.cm]{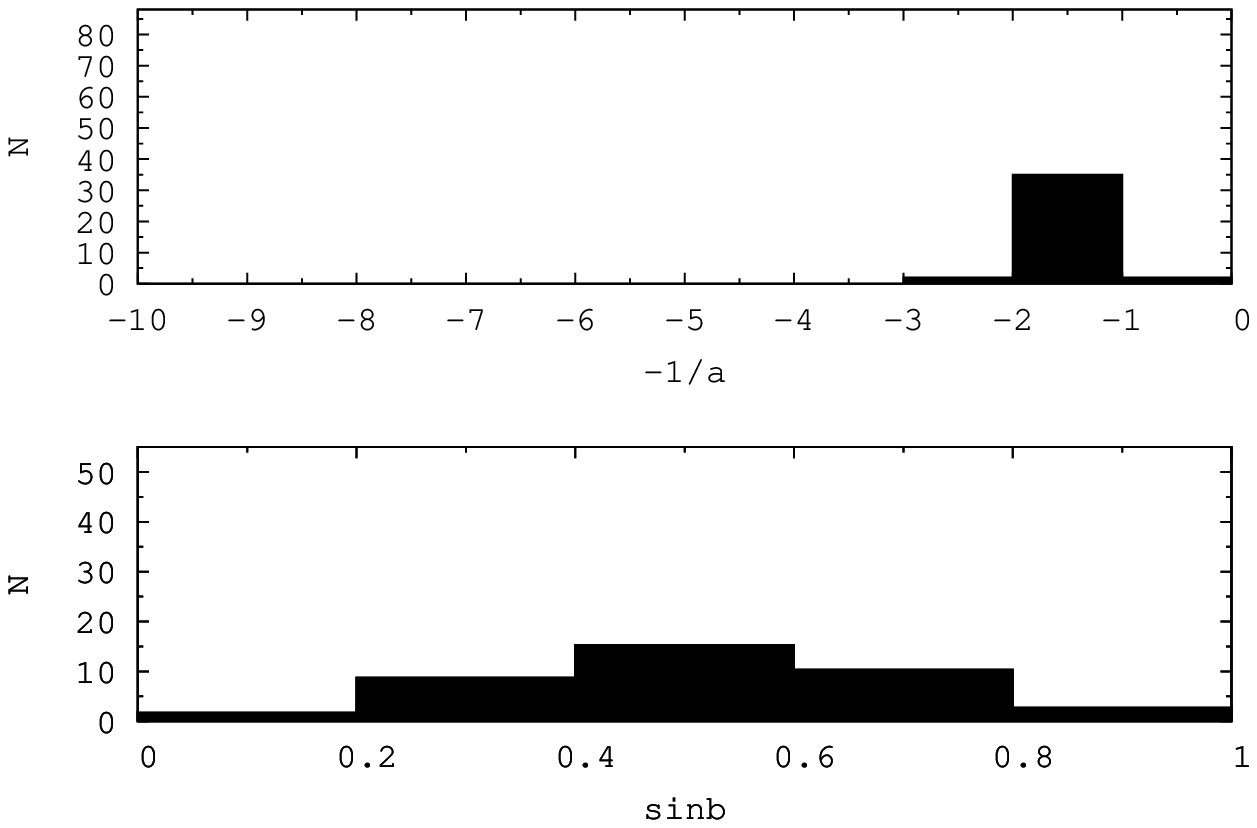}}
\end{picture}
\caption{Distributions of $-1/a$, where $a$ is the semi-major axis (top panels) and 
$|\sin b|$, where $b$ is the Galactic latitude of perihelion (bottom panels), for the comets 
entering the observable region during 170~Myr near the end of the simulation. When present, 
numbers in the top-left corners of $-1/a$ distribution panels correspond to comets 
with $-1/a<-1 \times 10^{-4}~{\rm AU}^{-1}$. The left column corresponds to the model
with Galactic tide alone, the middle column to passing stars alone, 
and the right column to the model with both effects.}
\label{fig:IN_4.55d9-4.72d9}
\end{center}
\end{figure}

Figure~\ref{fig:IN_4.55d9-4.72d9} shows the distributions of the opposite of the inverse 
semi-major axis 
$(-1/a)$ and the sine of the Galactic latitude of perihelion (for clarity we use the absolute 
value $|\sin b|$) of the comets entering the observable region, \ie, heliocentric distance smaller 
than 5~AU, during a typical 170~Myr interval near the end of our simulation, where no strong 
comet showers are registered. We show an average of three such periods, \ie, $4.38-4.55$~Gyr, 
$4.55-4.72$~Gyr and $4.80-4.97$~Gyr. In fact, comparing the three data sets, we find a rather 
good agreement, so that tentatively, the expected error of the mean is not very large. Three 
histograms are shown for each quantity: the one in black is for the model with Galactic 
tides only, the grey one is for the model with only stellar perturbations, 
and the white one shows the combined model.

After more than 4~Gyr the Galactic tides alone are practically only able to inject comets into 
the observable region if $a>50\,000$~AU, so that the non-integrable part of the tides may 
provide new comets into the emptied infeed trajectories of the vertical component. Thus the 
feeble flux of new observable comets is strictly confined to the outermost parts of the Oort 
cloud. If only the stellar perturbations are at work, the injected comets are almost as few 
as in the case of the Galactic tides. However, the distribution of $- 1/a$ shows that the stellar 
perturbations are relatively efficient injectors of comets with semi-major axes in the whole range 
from $25\,000$ to more than $100\,000$~AU, and there is some marginal infeed all the way into 
the inner core. Note that this concerns a time interval without any strong comet showers.

When both the processes are at work, the number of comets entering the observable zone is 206, 
about 86\% more than the sum of the two separate contributions ($39 + 72$). This estimate of $\tau$ is 
a bit higher than for the entire 1~Gyr interval, listed in Table~\ref{tab:tau}, because the 
three intervals have been selected as particularly calm, avoiding even the smaller peaks of $N_S$ 
that can be seen in Fig.~\ref{fig:flux_all}. We have shown above that larger values of $N_S$ 
lead to smaller values of $\tau$. The distribution of $ -1/a$ is as wide as for the stellar 
perturbations alone. However, the picture has changed, since the additional 86\% of the comets are 
strongly concentrated to the interval from $-4\times 10^{-5}$ to $-2\times 10^{-5}~{\rm AU}^{-1}$ 
($25\,000\,<\,a\,<\,50\,000$~AU). The local values of $\Delta N_C$ for the five $1/a$ intervals 
$(0-1)$, $(1-2)$, $(2-3)$, $(3-4)$ and $(4-5)\times10^{-5}$~AU$^{-1}$ are $-2$, $-22$, $+63$, $+36$ 
and $+10$, respectively. We will comment on the negative values of the first entries in connection 
with Fig.~\ref{fig:IN_6.3d8-8d8}.

We see that the mechanism of synergy that increases the flux of injections in the combined model 
prefers the range of semi-major axis ($a>30\,000$~AU) where the vertical Galactic tide is able 
to provide the injections, once the relevant trajectories are populated. But there is an 
important extension of the synergy to smaller semi-major axes as well, extending at least to 
$a\simeq20\,000$~AU. We conclude that both the above-described synergy mechanisms must be at 
work. The repopulation mechanism is obviously important, but the shift to smaller semi-major 
axes can only be explained by the `constructive interference' mechanism.

Looking at the distributions of $|\sin b|$, indeed the signature of the Galactic tide is clearly 
present in the left diagram and absent in the middle one. However, it appears again to some 
extent in the right-hand diagram, where the combined model is presented. Thus we have evidence 
that the synergetic injection of comets in the combined model carries an imprint in the latitudes 
of perihelia similar to that of the Galactic tide, though the feature is strongly subdued. In fact, 
while the subdued tidal imprint is consistent with an important role being played by the 
`constructive interference' synergy mechanism, our combined model does not appear to reproduce 
the observed $|\sin b|$ distribution of new Oort Cloud comets. An in-depth study of this problem 
and a consideration of ways out of this possible dilemma are deferred to future papers.

The shaping of the $b$ distribution by the Galactic tide was first simulated numerically for a 
realistic Oort Cloud model by \inlinecite{MAT.WHI:89}. However, the left panels of 
Fig.~\ref{fig:IN_4.55d9-4.72d9} show a behaviour that is in stark contrast to their results. 
Practically all our tidal injections occur for $a>50\,000$~AU, where \inlinecite{MAT.WHI:89} found 
no tidal imprint in the $b$ distribution because of complete loss cone filling independent of $b$. 
In the light of our results this can be seen as an artifact of their assumption of complete 
randomization of the Oort Cloud orbit distribution. Indeed, as we shall find below 
(Table~\ref{tab:flc_gsc}), the tidal loss cone filling for $50\,000<a<100\,000$~AU towards the end 
of our simulation is far from complete, and therefore we see the imprint of the tide in our $b$ 
distribution.

\begin{figure}[ht!]
\begin{center}
\setlength{\unitlength}{1.cm}
\psfrag{-1/a}[cc]{\tiny $-1/a\times 10^5~{\rm (AU^{-1})}$}
\psfrag{N}{\tiny $$}
\psfrag{sinb}{\tiny $|\sin b|$}
%\psfrag{-1}{}
\begin{picture}(11.8,6)
\put(7.56,0.){\includegraphics[width=4.cm,height=6.cm]{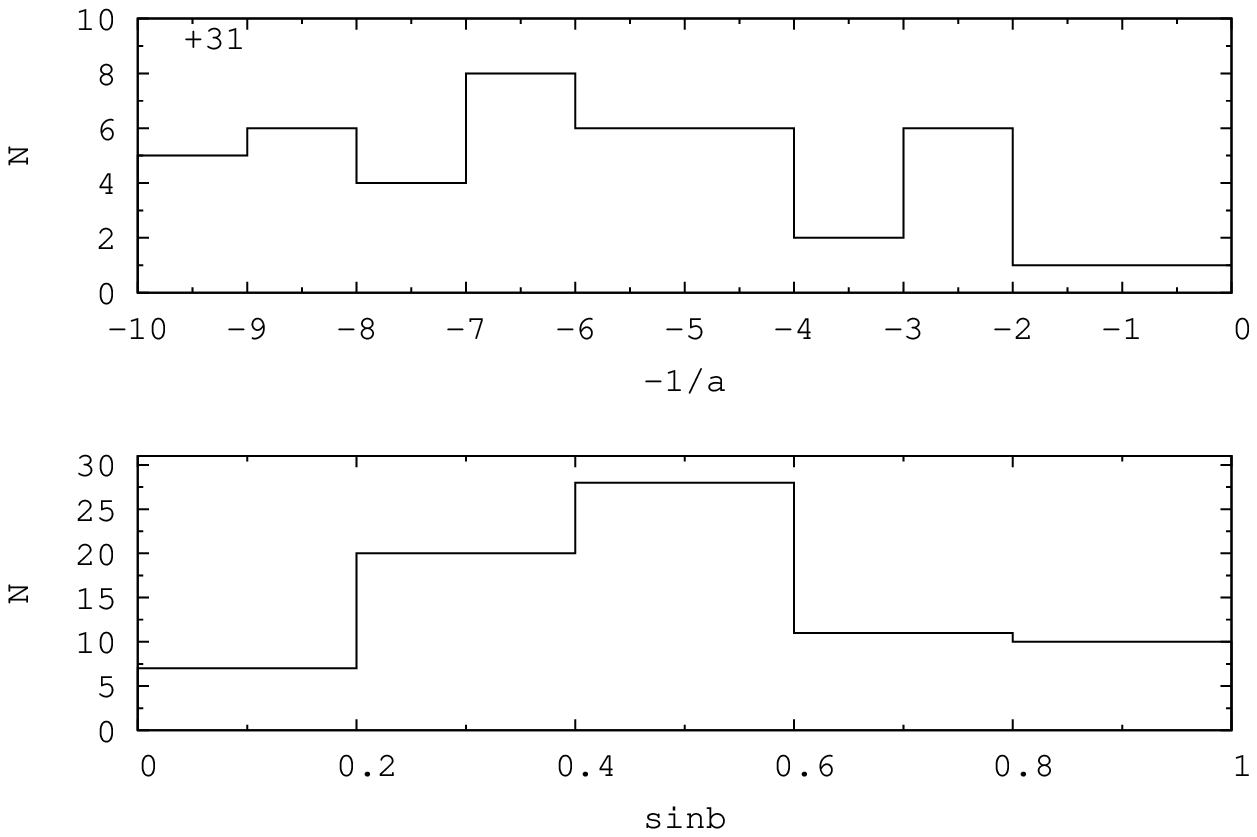}}
\put(3.80,0.){\includegraphics[width=4.cm,height=6.cm]{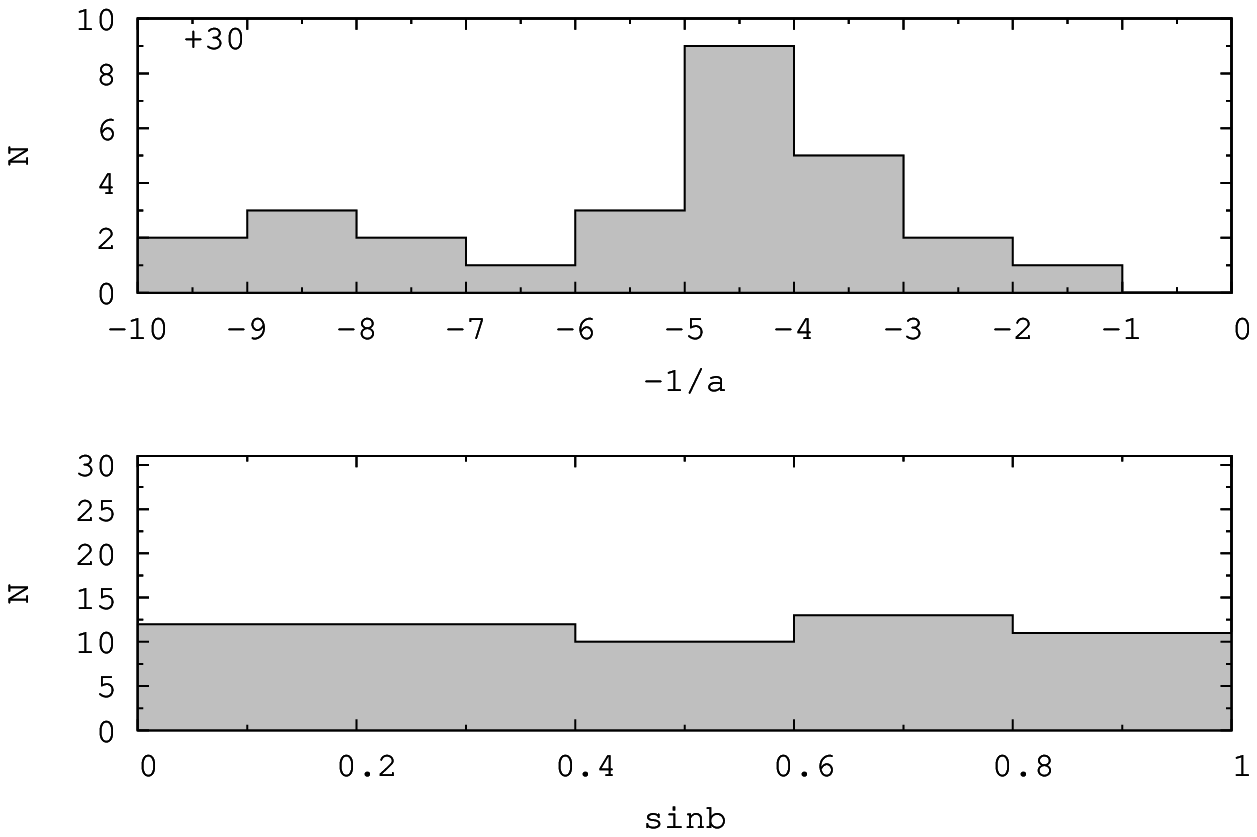}}
\psfrag{N}{\tiny $N$}
\put(0,0.){\includegraphics[width=4.cm,height=6.cm]{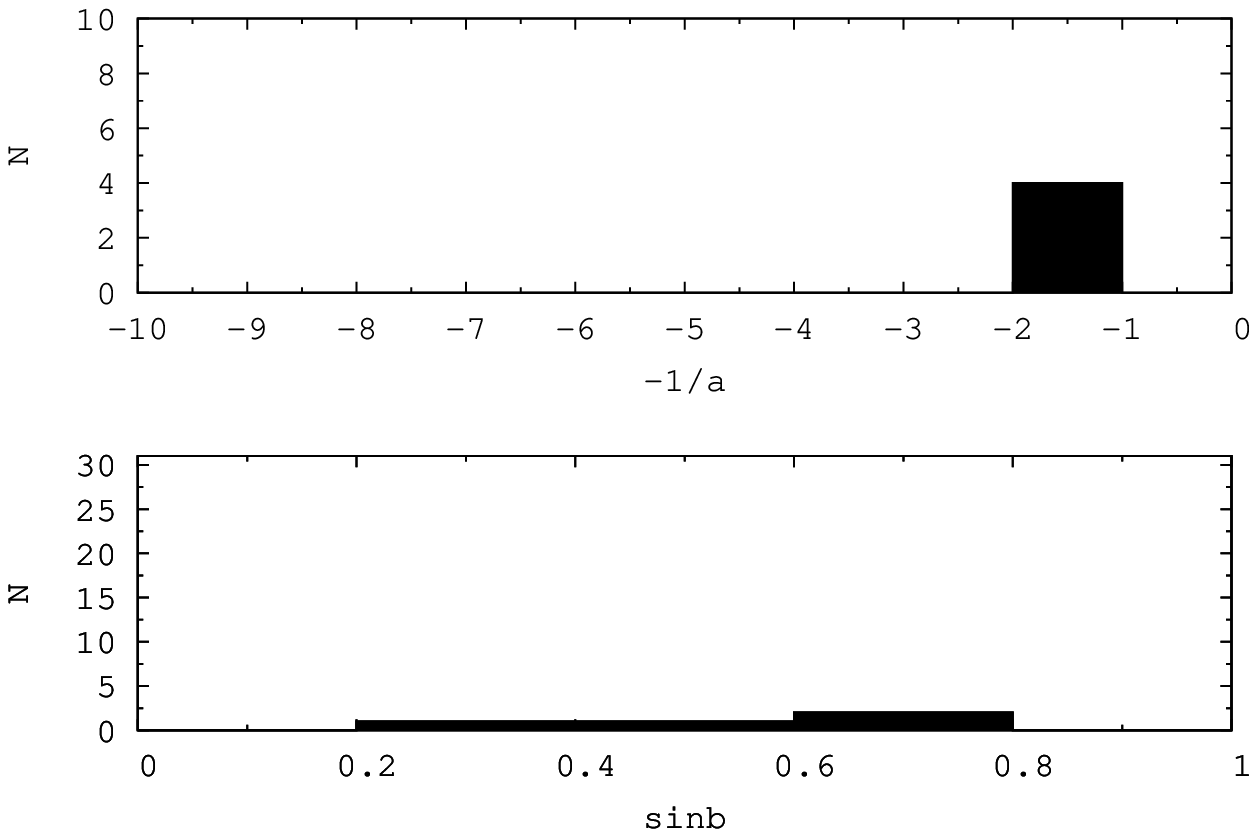}}
\end{picture}
\caption{Same as Fig.~\ref{fig:IN_4.55d9-4.72d9} but considering the comets that enter the 
observable region during a shower between $3.85$ and $3.86$~Gyr. The shower is due to a M5 star 
with impact parameter $2\,055$~AU, velocity 17.7~km/s and mass $0.21~M_\odot$.}
\label{fig:IN_3.85d9-3.86d9}
\end{center}
\end{figure}

\begin{figure}[ht!]
\begin{center}
\setlength{\unitlength}{1.cm}
\psfrag{-1/a}[cc]{\tiny $-1/a\times 10^5~{\rm (AU^{-1})}$}
\psfrag{N}{\tiny $$}
\psfrag{sinb}{\tiny $|\sin b|$}
%\psfrag{-1}{}
\begin{picture}(11.8,6)
\put(7.56,0.){\includegraphics[width=4.cm,height=6cm]{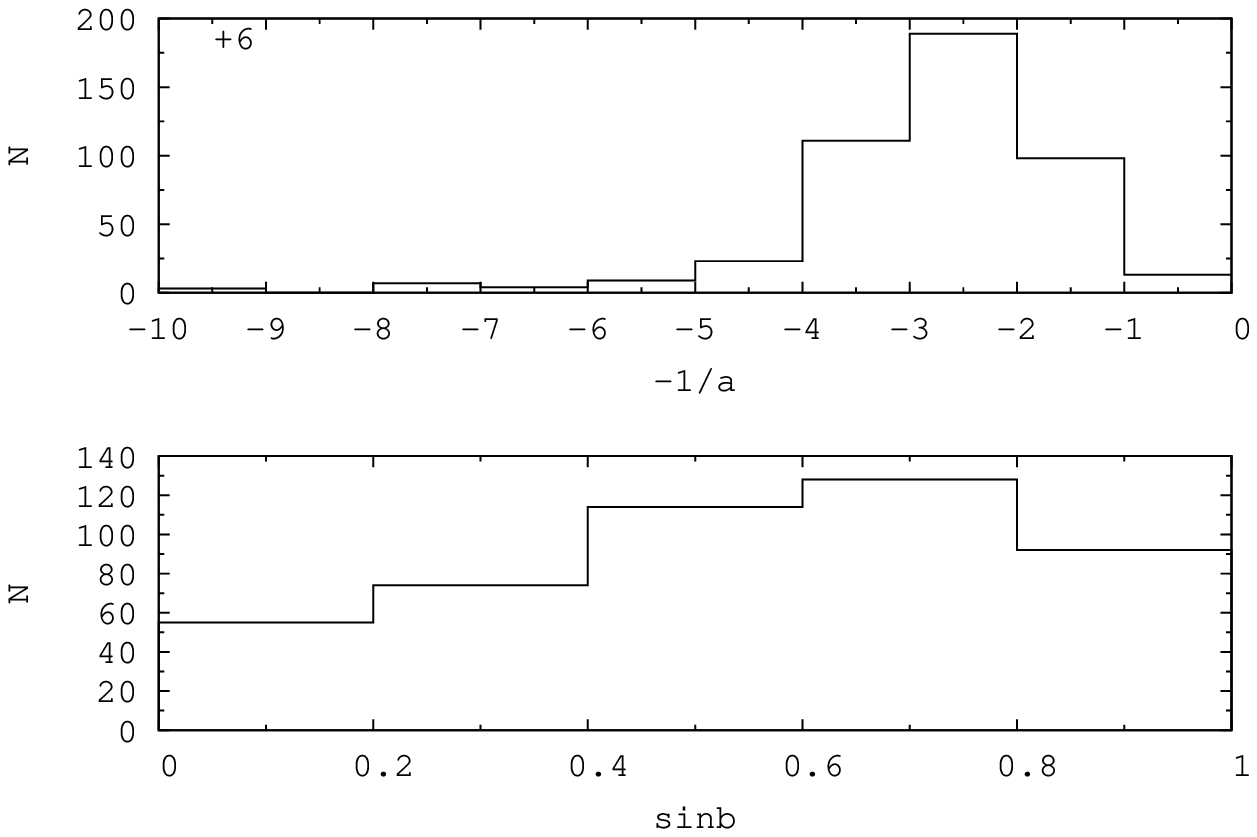}}
\put(3.80,0.){\includegraphics[width=4.cm,height=6cm]{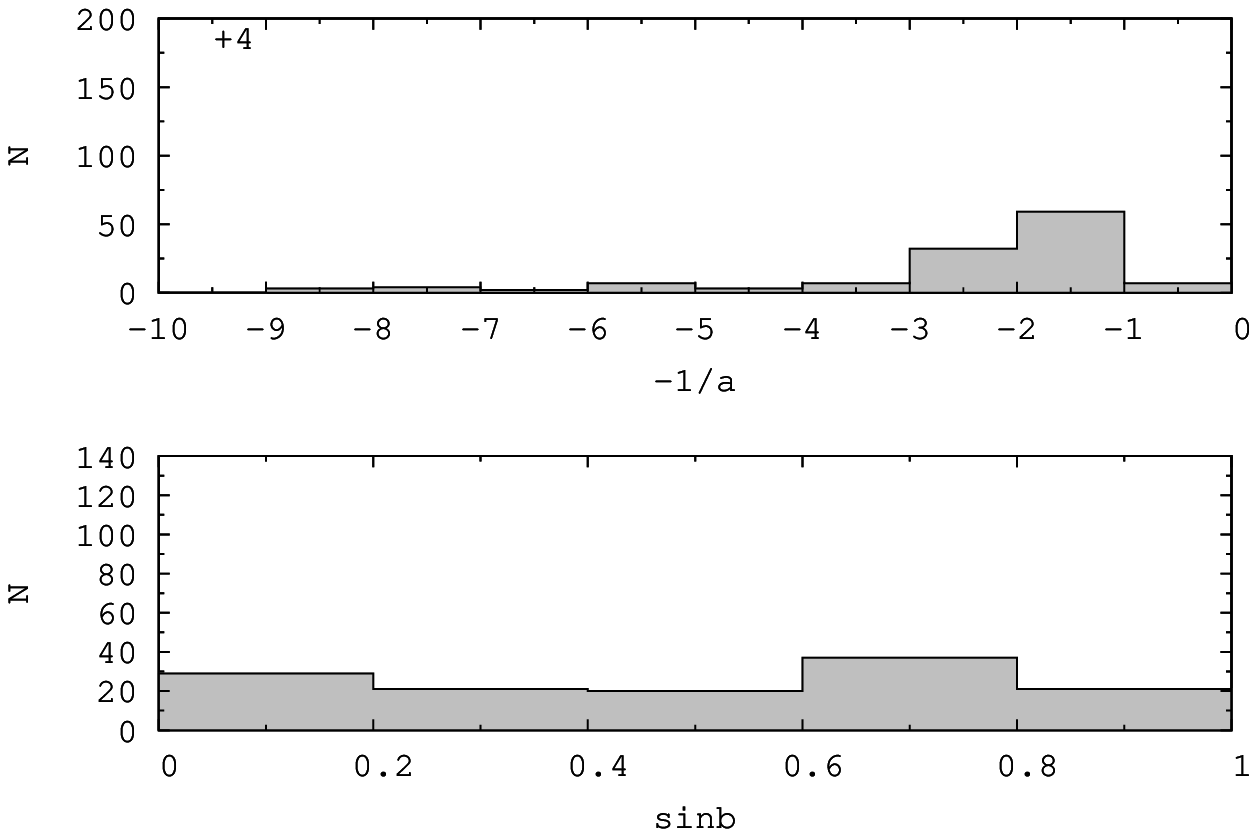}}
\psfrag{N}{\tiny $N$}
\put(0.,0.){\includegraphics[width=4.cm,height=6cm]{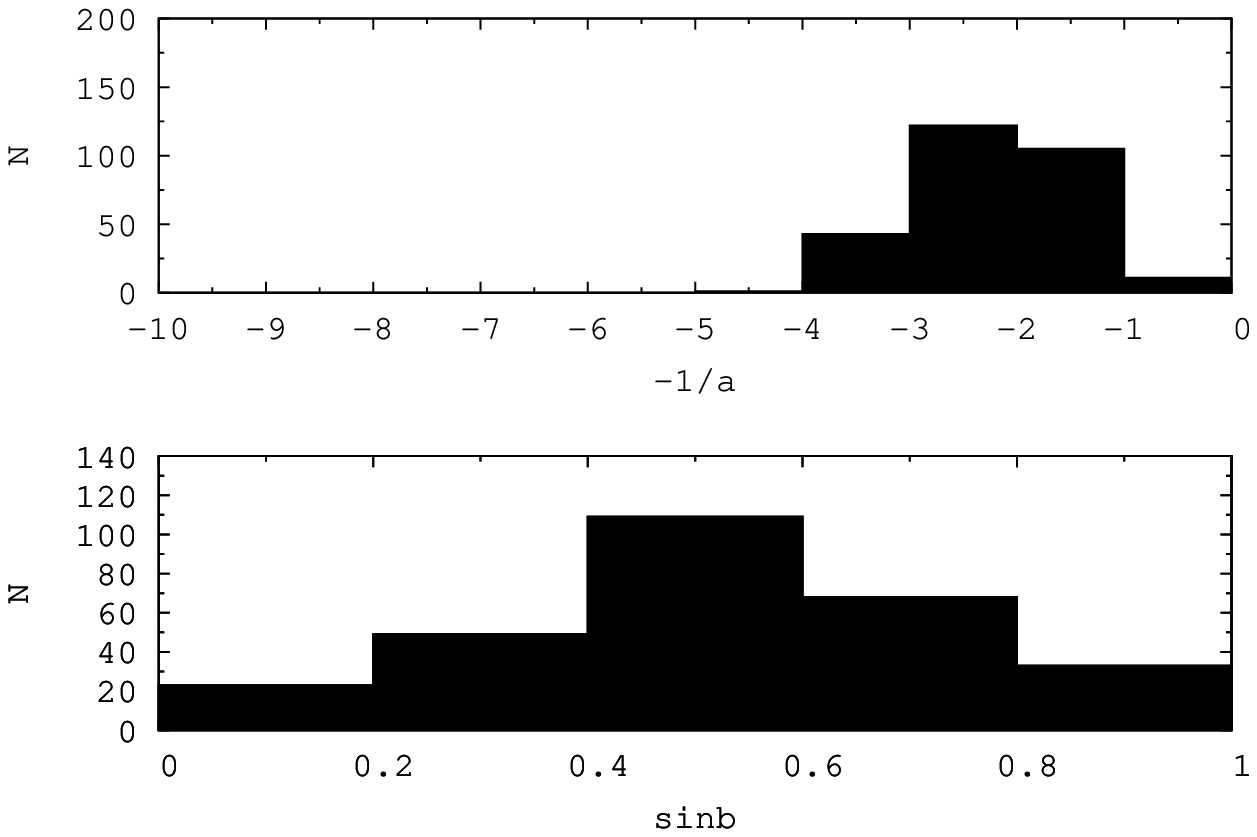}}
\end{picture}
\caption{Same as Fig.~\ref{fig:IN_4.55d9-4.72d9} but considering the comets entering the 
observable region during 170~Myr near the beginning of the simulation.}\label{fig:IN_6.3d8-8d8}
\end{center}
\end{figure}

In Fig.~\ref{fig:IN_3.85d9-3.86d9} we show the corresponding distribution of $-1/a$ and $|\sin b|$ 
for the 10~Myr interval from $3.85$ to $3.86$~Gyr, where Fig.~\ref{fig:flux_10d6} shows that the 
number of observable comets has a high peak due to a moderately strong shower. Occurring near the 
middle of the period, this shower dominates the time-integrated injection rate. The trigger is a M5 
star with an impact parameter $d\simeq2\,000$~AU and a velocity $v_\star\simeq18$~km/s. 

The mid and right-hand $-1/a$ distributions show that, as soon as stars are involved, the 
injection of comets now extends over the whole cloud, including an important fraction from 
the inner core with $a\,<\,10\,000$~AU. In fact, the synergy effect is now very strong in 
the range from $10\,000$ to $20\,000$~AU, amounting to $\tau>150$\%. This is unexpected 
on the basis of both the above-mentioned mechanisms, since we are discussing orbits too 
far inside the tidal injection limit. We are instead led to hypothesize a different 
mechanism. In the present case we are comparing the number of comets injected by a particular, 
deeply penetrating star from the mentioned range of semi-major axes in the stars-only vs 
the combined model. In the absence of the Galactic tides it is likely that orbits with 
perihelia close to but outside the loss cone have been depleted by the preceding cometary 
showers, while in the combined model the disk tide provides a regular transfer of comets 
into this zone on a Gyr time scale, thus compensating for the losses. This means that the 
synergy works in the opposite sense compared to the normal situation outside the showers. 
{\it During a shower the tides are providing the material for injections by the stars, while 
in the normal situation the stars are providing the material for tidal injections.} The 
absence of a synergy in the inner core may be explained by the very long time scale for 
tidal torquing, or by a smaller degree of depletion of the source region for stellar 
injections.

Naturally, in the stars-only model the shower does not carry any signature in the 
distribution of $|\sin b|$. The combined model does not exhibit any significant signature 
either, but there may nonetheless be a slight tendency. In case this is real, it might 
possibly reveal a somewhat more efficient synergy in the $10\,000-20\,000$~AU range for 
the orbits experiencing a faster tidal decrease of $q$.

Let us now consider the situation at the beginning of the simulation, before the tides
have had the time to completely empty the tidal infeed trajectories in the outer part of the 
cloud. The results are shown in Fig.~\ref{fig:IN_6.3d8-8d8} for a period between $0.63$ 
and $0.80$~Gyr, when no strong showers are noted. The number of comets entering into the 
observable region is 282, 128 and 463 for the tides, the stars, and the combined model, 
respectively. The action of the tides is still quite strong, since the infeed trajectories 
in the interval $20\,000<a<50\,000$~AU are not yet severely depleted (cf. Fig.~\ref{fig:dqt}). 
Therefore the net synergy effect amounts to only $13$\%.

The local $\Delta N_C$ values for the same five $1/a$ intervals as we discussed in 
connection with Fig.~\ref{fig:IN_4.55d9-4.72d9} are in this case $-5$, $-66$, $+35$, $+61$ 
and $+19$. We do not see any significant synergy effect for more tightly bound orbits. 
The distribution of positive synergy over the $2-5\times10^{-5}$~AU$^{-1}$ range is 
similar to that of Fig.~\ref{fig:IN_4.55d9-4.72d9}, and our conclusions about the relevance of 
the two mechanisms are the same. Note that in both cases we see negative $\Delta N_C$ values 
in the two outermost $1/a$ ranges ($0-2\times10^{-5}$~AU$^{-1}$). The fundamental reason is the 
one discussed above in order to explain the negative $\tau$ values in the very beginning of the 
simulation, \ie, a saturation effect of the loss cone when both injection effects individually 
are able to cause a large degree of filling.
The distributions of $|\sin b|$ exhibit the same features as in Fig.~\ref{fig:IN_4.55d9-4.72d9} 
and lead to the same conclusion: when both tides and stars act in synergy, the signature 
of the Galactic tide may be seen but appears quite marginal. 

Since we noted in Sect.~4 that the replenishment of the tidal infeed trajectories by stellar 
perturbations is not complete during the later part of our simulation, an obvious consequence
is that the filling of the loss cone for the relevant semi-major axes cannot be complete either. 
To quantify this statement, we consider the rate of perihelion passages $\dot{n}$ with $q<q_0$ 
as a function of the semi-major axis, assuming complete loss cone filling and a completely 
thermalized Oort cloud \cite{HIL:81,BAI.STA:88}:
\begin{equation}\label{eq:loss_cone}
\dot{n}(a)=N_{\rm OC} \cdot f(a) \cdot \frac{2\,q_0}{a}\cdot a^{-3/2}.
\end{equation}
Here, $N_{\rm OC}$ is the number of comets in the entire Oort cloud (initially $10^6$ in our model)
and $f(a)$ the frequency function describing the distribution of semi-major axes: 
$f(a) \propto a^{-1.5}$ initially in our model.

We have computed injection rates in the combined model for the three time intervals of 
Figs.~\ref{fig:IN_4.55d9-4.72d9}--\ref{fig:IN_6.3d8-8d8} using Eq.~(\ref{eq:loss_cone}) and 
finding the integrals $\int N_{\rm OC} f(a) a^{-5/2} da$ over different ranges of $-1/a$ 
directly from the simulation output at neighbouring moments. The calculation of these 
integrals is done by simply adding the values of $a^{-5/2}$ of all the comets found in the 
relevant range. Multiplying by the length of each interval and putting $q_0\,=\,5$~AU, 
we find the numbers $N_{\rm comp}$ listed in Table~\ref{tab:flc}.

%\vspace{.5cm}

\begin{table}[h!]
\centering
%\begin{landscape} 
{\tiny
\begin{tabular}{|c|ccc|ccc|ccc|}
$\Delta (1/a)$ & \multicolumn{3}{c}{Beginning}          & \multicolumn{3}{c}{End, showers}          & \multicolumn{3}{c|}{End, quiescent}  \\
\cline{2-10}
$(10^{-5}~{\rm AU}^{-1})$  & $N_{\rm comp}$ &  $N_{\rm sim}$ & $f_{\rm lc}$ &  $N_{\rm comp}$ &  $N_{\rm sim}$ & $f_{\rm lc}$ &  $N_{\rm comp}$ &  $N_{\rm sim}$ & $f_{\rm lc}$  \\
\hline
$(0-2) $ & 106    &   111        & $\sim100$\% &   3.2   &    2    & $\sim60$\%  &      48     & 40  & 83\% \\ 
$(2-3) $ & 303    &   189        &   62\%  &      10   &    6    & $\sim60$\%  &     160     & 83  & 52\% \\ 
$(3-4) $ & 616    &   111        &   18\%  &      24   &    2    & $\sim8$\%   &     367     & 46  & 13\% \\ 
$(4-5) $ &1044    &   23         &   2.2\% &      44   &    6    &  14\%     &     692     & 14  & 2.0\% \\
$(5-10)$ &15600   &   23         &  0.15\% &     740   &   29    &   3.9\%   &   12100     & 13  & 0.11\% \\
$ >10  $ &672000  &    6         & 0.0009\% &  37200   &   31    &  0.08\%   &  626000     & 10  & 0.0016\% \\
\hline
\end{tabular}
}
\caption{Numbers of comet injections during the time intervals of 
Figs.~\ref{fig:IN_4.55d9-4.72d9}--\ref{fig:IN_6.3d8-8d8} for different ranges of inverse 
semi-major axis, as computed from Eq.~(\ref{eq:loss_cone}) and found from our simulation of the 
combined model. 
The ratio of simulated to computed number, expressed in percent, is also listed in each case.}
\label{tab:flc}
%\end{landscape}
%\end{center}
\end{table}

The numbers of comet injections for each $1/a$ range and each time interval can be read off from the 
figures, and they are listed as $N_{\rm sim}$ in the Table along with the ratios 
$N_{\rm sim}/N_{\rm comp}$, which give the filling factor of the observable part of the loss cone 
($f_{\rm lc}$). We 
find that this factor is close to 100\% in the beginning of the simulation for $a>50\,000$~AU 
and remains $>80$\% for such semi-major axes even towards the end during quiescent periods. 
But the factor drops rapidly with decreasing $a$ to values near 2\% at $a\simeq20\,000$~AU. 
These results may be compared with those of \inlinecite{HEI:90}, who used a similar procedure 
for deriving $f_{\rm lc}$. She did not consider semi-major axes $a>40\,000$~AU, and inside this limit 
we find somewhat smaller filling factors than she did, consistent with the fact that we use a lower value 
for the Galactic mid-plane density and somewhat higher stellar velocities.

Note that the filling factors have decreased somewhat, when we compare the final quiescent 
period with the initial one. Except in the outermost parts of the Oort Cloud, there is always 
a depletion of comets in the regions of phase space near the tidal infeed trajectories and in 
the vicinity of the loss cone, and this depletion grows slowly with time.

The numbers $N_{\rm sim}$ found for the shower period are too small to be statistically 
useful for the outer parts of the cloud, and the filling factors listed are very uncertain. 
However, we see an obvious effect in the inner parts, when comparing $f_{\rm lc}$ with the corresponding 
values of quiescent periods. The shower increases $f_{\rm lc}$ by factors $\sim20-100$, and 
thus the overall flux exhibits the peak seen in Fig.~\ref{fig:flux_10d6} due to comets with 
$a<20\,000$~AU.

\begin{table}
{\tiny
\begin{tabular}{|c|ccc|ccc|}
$\Delta (1/a)$ & \multicolumn{3}{c}{$f_{\rm lc}$ (Beginning)} & \multicolumn{3}{c|}{$f_{\rm lc}$ (End)} \\
\cline{2-7}
$(10^{-5}~{\rm AU}^{-1})$  & Tidal &  Stellar & Combined &  Tidal &  Stellar & Combined  \\
\hline
$(0-1) $ & $\sim400$\% & $\sim100$\% & $\sim200$\% & $\sim100$\% & $\sim100$\% & $\sim100$\% \\ 
$(1-2) $ & 86\%        & 60\%        & $\sim100$\% & 30\%        & 45\%        & 78\%        \\ 
$(2-3) $ & 36\%        & 10\%        &   62\%      & 0.6\%       & 10\%        & 52\%        \\ 
$(3-4) $ & 6.5\%       & 1.1\%       &   18\%      & --          & 2.3\%       & 13\%        \\ 
$(4-5) $ & 0.09\%      & 0.3\%       &   2.2\%     & --          & 0.5\%       & 2.0\%       \\
$(5-10)$ & --          & 0.1\%       &  0.15\%     & --          & 0.06\%      & 0.11\%      \\
$ >10  $ & --          & 0.0006\%    & 0.0009\%    & --          & 0.0008\%    & 0.0016\%    \\
\hline
\end{tabular}
}
\caption{Filling factors for the observable part of the loss cone, computed for different ranges of 
semi-major axis and separately for the three dynamical models (tides-only, stars-only, and combined).
}\label{tab:flc_gsc}
\end{table}

We have already made the remark that neither $\Delta N_C$ nor $\tau$ provides a fully satisfactory 
measure of the synergy effect, because they do not account for the difference of the number of Oort 
Cloud comets between different dynamical models -- especially towards the end of the simulation. 
After 5~Gyr the total number of comets in the combined model is only $\sim80$\% of that in the 
tides-only model, and if we concentrate on comets with $50\,000<a<100\,000$~AU where the losses 
are the largest, the ratio of the two models is only 35\%. In order to compensate for such effects 
we have computed the $f_{\rm lc}$ parameter separately for the three models and for all the ranges 
of $1/a$, and we present the results in Table~\ref{tab:flc_gsc}. The time periods referred to are 
the quiescent periods of Figs.~\ref{fig:IN_4.55d9-4.72d9} and \ref{fig:IN_6.3d8-8d8}.

The outermost energy range is empty in all models, when the simulation starts, but it gets populated 
quickly -- at least when stars are involved. We interpret the very large value of $f_{\rm lc}$ in 
the tides-only model at the beginning as evidence that the radial tide has not extracted comets into 
this energy range in a uniform manner, so that our assumption of thermalization when deriving 
$N_{\rm comp}$ is not justified. To a lesser extent this appears to be true also in the combined 
model, where stars have extracted many more comets. It is likely that this extraction too -- at 
the early time in question -- has not populated all the angular momenta in a thermalized fashion. 
However, the statistics is too poor to be confident about such conclusions. In any case, the loss 
cone filling is extremely efficient for all models, thus explaining the negative values of $\Delta N_C$.

For the next energy range we see the 
saturation effect again, especially in the beginning. At the end, the value of $f_{\rm lc}$ in the 
combined model is close to the sum of those in the other two models. Hence there is no apparent 
synergy in this case, but probably a real synergy has been concealed by the saturation effect. In 
any case the large negative value of $\Delta N_C$ results entirely from the small number of comets 
in the combined model, as discussed above. In the next three energy ranges ($20\,000<a<50\,000$~AU) 
we see that $f_{\rm lc}$ in the combined model is much larger than the sum of the two other entries, 
and for $a<20\,000$~AU the effect continues: adding the tides to the stars increases the loss cone 
filling by a factor $1.5-2$.

\section{Discussion and conclusions}\label{sec:conclu}

We have simulated the evolution of the Oort cloud over 5~Gyr, using for initial conditions a relaxed 
model with a distribution of semi-major axis $f(a) \propto a^{-1.5}$ within the interval 
$3\,000 - 100\,000$~AU. This model is based on the results of simulation of Oort Cloud 
formation and evolution by \inlinecite{DUNetal:87}. However, we do not find this to be a steady 
distribution. More comets are lost from the outer parts of the cloud than can be replaced from inside, 
so that our model cloud evolves into a distribution close to $f(a) \propto a^{-2}$ -- \ie, flat in 
$1/a$.

Our dynamical model has two main limitations. We do not treat encounters with very massive Galactic 
perturbers, such as star clusters or Giant Molecular Cloud complexes, the justification being that they 
occur so rarely that the current Solar System is unlikely to feel the direct reverberations of any such 
encounter, and that even if they modify the structure of the Oort cloud, our interest is not primarily 
in its dynamical history but rather in the way stars and Galactic tides currently interact when injecting 
observable comets.

Moreover, we do not treat planetary perturbations in any direct manner. Like most previous investigators 
({\it e.g.}, \opencite{HEI:90}) we use a loss cone defined by a limiting perihelion distance (in our 
case, 15~AU) outside which no planetary effects are included and inside which all comets are 
considered lost from the cloud through perturbations by Jupiter and Saturn. In terms of ``transparency'' 
of the planetary system \cite{DYB.PRE:97,DYB:05}, our model is completely opaque ($P=1$). This 
means that we are limiting 
our attention to a subset of the observed population of ``new'' Oort cloud comets, \ie, those that have 
jumped directly from $q\,>\,15$~AU into their observed orbits with $q\,<\,5$~AU. We are neglecting the 
rest of the population, which consists of comets that passed perihelia in the outer part of the loss cone 
without being perturbed away before arriving into observable orbits. We are also neglecting a possible 
contribution to the observed new comets by a ``leakage'' from the scattered disk \cite{LEVetal:06}. 
Therefore we prefer not to draw any conclusions in this paper regarding the total number of Oort Cloud
comets or the exact values of the filling factors. Nor do we claim to make any prediction on the 
detailed shape of the $1/a$ distribution of new 
Oort Cloud comets, until we have included the planetary perturbations in a realistic manner.

We have shown that the concept of tidal and stellar torquing time scales \cite{DUNetal:87} gives a very 
incomplete picture of the speed of comet injection, whether it may concern Galactic tides or stellar 
encounters. The distribution of injection times is largely shaped by other effects -- like comet showers 
or the repopulation of the emptied infeed trajectories of the disk tide due to the non-integrable part 
of the tides or stellar perturbations.

We have also shown how -- for semi-major axes large enough for the tide to populate observable orbits 
-- the regions of the phase space occupied by trajectories leading into the loss cone get depleted 
during the first Gyr of Oort cloud evolution. This would leave little chance for the tide to produce a 
significant number of observable comets at the current time, were it not for the capability of stellar 
perturbations to replenish the tidal infeed trajectories.

We have indeed demonstrated that, during the later parts of our simulation, there is a very important 
synergy effect of the Galactic tide and stellar perturbations such that the combined injection rate is 
on the average $\sim70$\% larger than that of the stars alone plus that of the tide alone. This synergy 
is strongest for semi-major axes between $\sim20\,000-50\,000$~AU but continues all the way 
into the inner core. During comet showers the synergy effect in the outer parts of the cloud practically 
disappears, but the one affecting the inner parts becomes very important. 

We have identified two mechanisms for the synergy during quiescent periods in the outer parts of the 
Oort Cloud. One is that the stellar perturbations provide a supply of new comets that replenishes the 
depleted tidal infeed trajectories, and the other is that the gain of comet injections, when stellar 
perturbations decrease the perihelion distance, dominates over the loss caused by opposing perturbations. 
For the synergy of the inner cloud we hypothesize that the Galactic tides provide the material for 
stellar injections by slowly feeding the region of phase space in the vicinity of the loss cone. Thus, 
the general picture spawned by our results is that injection of comets from the Oort Cloud is 
essentially to be seen as a team work involving both tides and stars. It appears meaningless to rank 
the two effects in terms of strength or efficiency.

Indeed, for the smaller semi-major axes the Galactic tide does not dominate the injection of comets, 
contrary to the conclusions of \inlinecite{HEIetal:87} and \inlinecite{HEI:90}.\footnote{The main reason 
for this discrepancy is that the Heisler papers considered injections into the loss cone -- 
mainly by slight perturbations of $q$ across the limiting value $q_c\,=\,10$~AU -- while we consider large 
jumps from $q\,>\,15$~AU into the observable region with $q\,<\,5$~AU. Interestingly, 
\inlinecite{HEIetal:91} commented that the injection into orbits with $a \, \lta \, 20\,000$~AU and 
$q\,<\,2$~AU is indeed dominated by stellar perturbations.} It only contributes to a synergy with 
stellar perturbations, and without the stars one would not have any injections of comets 
with $a\,\lta\,20\,000$~AU. 

The distribution of Galactic latitudes of perihelia of the observable comets exhibits a maximum for 
$|\sin b|\simeq0.5$ as expected in the tides-only model, but in the combined model this feature can 
hardly be seen at all. The tides form part of the synergetic injection, but their imprint is largely 
washed out by the stellar contribution. But, likely due to the role of the tides in helping the stars 
to create comet showers, the pattern can be seen at least as clearly during a shower as during the 
quiescent periods. Therefore, it tentatively appears that the shape of the observed $b$ distribution 
can not be used to indicate whether we are experiencing any shower at present. However, since none of 
our model distributions appears to agree with the observed one, we have to defer any conclusions 
to future papers. It may be interesting to see, for instance, if the leakage from the scattered disk 
into the Oort Cloud with an ensuing direct transfer into observable new comets may alleviate the 
problem.
 
We have measured the filling of the observable part of the loss cone by comparing our simulated 
injection rates for different intervals of semi-major axis with the rates of observable perihelion 
passages ($q\,<\,5$~AU) computed for a completely thermalized distribution of cometary orbits 
involving a filled loss cone. The deficiency of our simulated rate likely reflects not only a lack 
of comets in the loss cone but a general depletion in a wider phase space region in its vicinity, 
as remarked by \inlinecite{HEI:90}. Our results can be compared with hers, and in contrast to her 
inference that $f_{\rm lc}$ may level out at $\sim60$\% for $a\gta30\,000$~AU, we find an average 
filling factor during quiescent periods in the current 
Solar System, which drops steadily from $\sim100$\% at $a>100\,000$~AU to 1\% or less at $a<20\,000$~AU 
in the combined model. However, there are important differences between the two investigations, one 
being that she simulated a much shorter time period than we do, and in addition our parameters for the 
Galactic tides and stellar encounters also differ from hers.

In agreement with \inlinecite{WEI.HUT:86}, we find that cometary showers do not significantly increase 
the loss cone filling at large semi-major axes. However, near $25\,000$~AU there is an abrupt change 
into the regime of the inner cloud, where the filling factor increases by orders of magnitude during 
moderate to strong events. The showers of course involve direct injections by the passing stars, but 
the synergy with the Galactic tide is as important as during quiescent periods.

\section*{Acknowledgments}
We are greatly indebted to the referees of this paper, Ramon Brasser and Julio A.~Fern\'andez, for 
having pointed out various weaknesses in a previous version, which inspired us to exert more care 
in all parts of our analysis. M.F. is grateful to GDRE 224 CNRS INdAM, GREFI-MEFI for financial support.
For H.R., this work was supported by Grants nr. 78/06 and 119/07:1 of the Swedish National Space Board.
The work of G.B.V. was supported by the contract ASI/INAF I/015/07/0.

%%%%%%%%%%%%%%%%%%%%%%%%%%%%%%%%%%%%%%%%%%%%%%%%%%%%%%%%%%%%%%%%%%%%%%%%%%%%%%%%%%%%%%%%%%%%%%%%

%\bibliographystyle{klunamed}
%\bibliography{longarticle}

\label{lastpage}

\end{article}
\end{document}